\documentclass[fleqn,usenatbib]{mnras}
\usepackage{newtxtext,newtxmath}

\usepackage[T1]{fontenc}
\usepackage{ae,aecompl}
\usepackage{caption}
\usepackage{hyperref}
\usepackage{graphicx}	
\usepackage{amsmath}	
\usepackage{amssymb}	
\usepackage{mydefs}
\usepackage[dvipsnames]{xcolor}
\usepackage{threeparttable}
\usepackage{ulem}
\usepackage{rotating}
\usepackage[export]{adjustbox}
\usepackage{todonotes,ifthen}
\newcommand{\Nzbin}{N_{\rm zbin}}
\newcommand{\zph}{z_{\rm ph}}
\newcommand{\zspec}{z_{\rm spec}}

\title[BAO from kinematic weak lensing]{Detecting Baryon Acoustic Oscillations in Dark Matter from Kinematic Weak Lensing Surveys}
\author[Ding et al.]
{\parbox{\textwidth}{Zhejie Ding$^{1}$\thanks{E-mail: \texttt{zd585612@ohio.edu}}, Hee-Jong Seo$^{1}$\thanks{E-mail: \texttt{seoh@ohio.edu}}, Eric Huff$^{2}$, Shun Saito$^{3,4,5}$, Douglas Clowe$^{1}$}\vspace{0.4cm}\\
\parbox{\textwidth}{$^{1}$ Department of Physics and Astronomy, Ohio University, Clippinger Labs, Athens, OH 45701, USA\\
$^{2}$ Jet Propulsion Laboratory, California Institute of Technology,
4800 Oak Grove Drive, Pasadena, CA 91109, USA\\
$^{3}$ Max-Planck-Institut f\"{u}r Astrophysik, Karl-Schwarzschild-Star{\ss}e 1, D-85740 Garching bei M\"{u}nchen, Germany\\
$^{4}$ Department of Physics, Missouri University of Science and Technology, 1315 N. Pine St., Rolla MO 65409, USA\\
$^{5}$ Kavli Institute for the Physics and Mathematics of the Universe (WPI), The University of Tokyo Institutes for Advanced Study, The University of Tokyo, 5-1-5 Kashiwa-no-Ha, Kashiwa, Chiba 277-8583, Japan
}}

\date{Accepted XXX. Received YYY; in original form ZZZ} 

\pubyear{2018}

\begin{document}

\maketitle

\begin{abstract}
We investigate the feasibility of extracting Baryon Acoustic Oscillations (BAO) from cosmic shear tomography. 
We particularly focus on the BAO scale precision that can be achieved by future spectroscopy-based, kinematic weak lensing (KWL) surveys \citep[e.g.,][]{Huff13} in comparison to the traditional photometry-based weak lensing surveys. 
We simulate cosmic shear tomography data of such surveys with a few simple assumptions to focus on the BAO information, extract the spacial power spectrum, and constrain the recovered BAO feature. Due to the small shape noise and the shape of the lensing kernel, we find that a Dark Energy Task Force Stage IV version of such KWL survey can detect the BAO feature in dark matter by $3$-$\sigma$ and measure the BAO scale at the precision level of 4\% while it will be difficult to detect the feature in photometry-based weak lensing surveys. With a more optimistic assumption, a KWL-Stage IV could achieve a $\sim 2\%$ BAO scale measurement with $4.9$-$\sigma$ confidence. A built-in spectroscopic galaxy survey within such KWL survey will allow cross-correlation between galaxies and cosmic shear, which will tighten the constraint beyond the lower limit we present in this paper and therefore possibly allow a detection of the BAO scale bias between galaxies and dark matter.  
\end{abstract}


\section{Introduction}
Baryon acoustic oscillations are pressure waves in the early Universe that propagated in the hot plasma of photons and baryons that were tightly coupled via Compton scattering.  At the epoch of recombination, which is $\sim 300,000$ years after the Big Bang, the temperature of the Universe was low enough that electrons and protons combined to form hydrogens, decreasing the optical depth to Thompson scattering. Due to the photon decoupling, the propagating waves lost photon pressure, which dropped the sound speed, and the Universe was left with a frozen spherical overdensity shell around each random overdensity peak. Such frozen spherical overdensity waves are  imprinted in the distribution of cosmic microwave backgrounds as well as in the distribution of matter and galaxies in the later Universe. The largest distance that the sound wave had propagated before the epoch of recombination is called the sound horizon scale $r_s$ at recombination, which corresponds to about $150\Mpc$ today \citep{Hu_Sugiyama96, EH98}. 

The true physical scale of the sound horizon at recombination can be precisely measured from the cosmic microwave background radiation (CMB) data. The standard ruler test compares this physical scale with the scale of BAO imprinted in matter density distribution in the observational coordinates at low redshift. The metric between the physical coordinates and the observational coordinates such as right ascension, declination, and redshift encodes the expansion history of the Universe, which in turn depends on dark energy properties~\citep[e.g.][]{Weinberg13}. Since dark matter, the majority of matter, is not optically observable, we use galaxies as tracers to detect matter fluctuations, and therefore the BAO feature, as baryons and dark matter have fallen in the common gravity potential wells at low redshift. Since the first detection of BAO from galaxy surveys \citep{Eisenstein05, Cole05}, high precision BAO scale measurements have been obtained from various galaxy surveys, e.g. the 6dF Galaxy Survey \citep{Beutler11}, the WiggleZ Dark Energy Survey \citep{Blake11}, the Baryon Oscillation Spectroscopic Survey (BOSS) \citep{Alam16} and the extended Baryon Oscillation Spectroscopic Survey (eBOSS) \citep{Zhu18}, and will extend to greater precision and higher redshifts in upcoming surveys such as the Dark Energy Spectroscopic Instrument (DESI) \footnote{https://www.desi.lbl.gov} \citep{DESI_science}, the Prime Focus Spectrograph~\footnote{https://pfs.ipmu.jp}~\citep{PFS}, and the Euclid \footnote{https://www.euclid-ec.org}\citep{Euclid_whitepaper}.

However, galaxies are not perfect tracers of matter distribution. It has been shown that galaxy bias could bias the BAO scale relative to that of dark matter \citep[e.g.][]{SE05,PadBAOshift,SZ12,vlah2015, Senatore15,zv16, Blas16, Noda17}. In addition, galaxy clustering suffers distortion due to the peculiar velocity fields of the galaxies, which introduces additional bias on the BAO scale along the line of sight. Both effects on the BAO can, however, be substantially alleviated by the method called the density field reconstruction, \citep[e.g.][]{ESW07}. 

Another potential bias on the BAO scale from galaxy surveys has been recently suggested; the velocity field of baryons relative to dark matter was supersonic in the very early Universe. Baryons that were propagating in the form of sound waves would still be moving supersonically after recombination while they were settling down in the common gravitational well of matter. This supersonic streaming velocity would have prevented gas accretion and cooling that is essential for galaxy formation in dark matter halos. As a result, the galaxy distribution relative to the underlying dark matter would have been modulated by these effects, which can possibly generate a relative offset of the BAO scale in galaxy distribution to the one in dark matter \citep{Tseliakhovich10, Yoo11, Blazek16}. If one does not account such bias in the analysis, it will result in biased dark energy constraints from galaxy BAO surveys. 

Measuring the BAO scale directly from matter distribution can reveal the extent of such bias and therefore help us derive correct cosmological parameters.
Weak lensing (WL) has the advantage of detecting cosmological matter distribution directly. Images of distant galaxies are slightly distorted by the gravitational potential of the foreground large-scale structures as light passes by, which is referred as cosmic shear \citep[see the review, e.g.][]{Kilbinger15}. Since shear signal depends on the distribution of the intervening matter but not on its kinematics (i.e., the peculiar velocity field), the clustering derived from cosmic shear probes the geometry of space and the rate of structure growth without suffering systematics from galaxy bias or redshift-space distortions~\citep{Kaiser87}~\footnote{Here we do not consider observational systematics in weak lensing measurements such as point-spread-function (PSF), etc.}.
Therefore, it has become one of the main probes of the on-going and upcoming large-scale surveys, such as the Dark Energy Survey \citep[DES;][]{DES05}, the  Hyper Suprime-Cam survey \citep[HSC;][]{Subaru17}, the Large Synoptic Survey Telescope \citep[LSST;][]{Ivezic08}, the Wide-Field Infrared Survey Telescope \citep[WFIRST;][]{Spergel13a, Spergel13b}, and the Eulid \citep{Euclid_whitepaper}. 

Despite the advantage of directly probing matter clustering, it has been challenging to detect the BAO from weak lensing surveys~\citep[c.f.,][]{Grassi14}.
Shear signal is weak, only about $1\%$ of the intrinsic shape that itself is unknown. Noise in the shear signal due to the uncertainty in the intrinsic shape (due to the random intrinsic orientations or ellipticities) could be reduced by observing a large number of galaxy images. Observing more and more galaxies to fainter magnitudes (since the number of luminous galaxies is limited) sets a practical preference for imaging surveys to spectroscopic surveys. The large redshift error associated with imaging surveys then will further broaden the broad lensing efficiency kernel along the line of sight, which smears the clustering signal over a large range of distance along the line of sight. The broad lensing kernel causes mixing of a wide range of different physical scales that correspond to the same angular scales at different distances, making the extraction of a distinct feature such as BAO challenging~\citep[e.g.,][]{Simpson2006}.
Dividing galaxies into several tomographic bins and correlating statistics between bins \citep{Hu99} has been commonly used in WL survey analyses to partially resolve the line-of-sight information \citep[e.g.][]{Schrabback10,Hildebrandt17,Troxel17, SubaruY1}. However, the shape noise of shear signal increases as the galaxy number density decreases in tomographic bins, which limits the maximum amount of shear signal from tomography.

Recently, \citet{Huff13} (Huff13, hereafter) revived the idea of kinematic weak lensing (KWL, hereafter) that conducts WL using multi-object spectroscopy in combination with high-quality imaging data, based on the methods proposed in early literature \citep{Blain02, Morales06}. 
In this method, multi-object spectroscopy such as by the DESI spectrograph or the Prime Focus Spectrograph for the Subaru telescope is used to measure the kinematics (i.e., rotational velocity) of disk galaxies to distinguish the effect of shear from the effect of inclination. Shear oriented along the major axis of a rotating disk (i.e., the even parity component) is derived from the comparison between the observed ellipticity of the isophote and the inclination effect that is estimated based on the offset from the Tully-Fisher relation. The odd parity shear component skews the kinematic axes relative to the photometric axes and therefore is derived from the observed velocity along the sheared semi-minor axis of the disk galaxy. Using spectroscopy of course would limit the number of source galaxies we can observe while those selected would be brighter and well-resolved targets so that they are more robust to various weak lensing calibration biases~\citep[c.f.][]{Hirata2003} that we do not consider in this paper. There still remains a small error since, for example, even a face-on disk galaxy may not be round (see Huff13 for more details for the effective shape noise sources for KWL surveys), but this shape noise is estimated to be only 1/10 times that of the traditional shape noise, which corresponds to saving 100 times the typical number of galaxies required for WL. As a result, even after accounting for the substantially lower source galaxy number density achievable in spectroscopic KWL surveys, assuming one galaxy per ${\rm arcmin}^2$, Huff13 estimates that the net shape noise effect on the KWL clustering data can be 5-17 times (in $\sigma^2_{\epsilon}/n$ where $\sigma_\epsilon$ is the rms intrinsic shear and $n$ is the angular number density of source galaxies) less than the current and future photometry-only weak lensing (PWL, hereafter) surveys.
Also, by construction, the spectroscopic data can determine the cosmological redshift of each galaxy accurately and precisely (up to the peculiar velocity effects), without the additional convolution of the lensing kernel due to source redshift uncertainty along the line of sight. According to Huff13, a Dark Energy Task Force (DETF) Stage III version of such survey (covering 5,000 ${\rm deg}^2$) can return about 3 times more dark energy information than the DES in terms of figure of merit, and a DETF Stage IV version of such survey (covering 15,000 ${\rm deg}^2$) can return seven times more dark energy information than the DES.

In this paper, we test if we can directly detect the BAO feature in the matter distribution if such kinematic weak lensing surveys are available, taking advantage of the low shape and redshift noises associated with the KWL surveys. We note that this can be considered as an independent and additional gain to what was estimated in Huff13 where they focused on the broadband clustering signal by choosing a large redshift bin width for the tomographic bins. We also note that the intension of this paper is not to investigate the technical feasibility of such future survey, but rather to investigate the scientific advantage of such survey from the BAO perspective over a range of survey parameter choices.

In addition to being the first detection of BAO from the matter distribution, another importance of such measurement is that it will provide an additional and independent BAO measurement for testing the consistency  within the flat $\LCDM$ model between the Cosmic Microwave Background (CMB) data by the Planck mission~\citep{Planck15, Planck18} and the low redshift data. Currently, while the galaxy BAO measurements are consistent with the flat $\LCDM$ best fit of the CMB data~\citep[e.g,][]{Alam16}, weak lensing survey constraints~\citep[a combination parameter of the matter density $\Om$ and the amplitude of matter fluctuations $\sigma_8$ from e.g.,][]{Hildebrandt17,Hildebrandt18, KIDS,Joudaki17,Troxel17,SubaruY1} and the distance-ladder Hubble constant measurements~\citep[e.g.,][]{Riess18} show a $2-3\sigma$ level of inconsistency with respect to the Planck constraints with the flat $\LCDM$ assumption. A similar test with the matter BAO scale from KWL surveys will be crucial for understanding such consistency or tension between different probes, especially since the same data set will simultaneously provide the conventional weak lensing constraint on $\Om-\sigma_8$.

In this paper, we simulate auto and cross cosmic shear angular power spectra between tomographic redshift bins of KWL surveys as well as PWL surveys with a few simplifications in order to focus on the BAO information. We deconvolve the lensing efficiency kernel and derive the spacial power spectrum that corresponds to the maximum likelihood. We isolate the BAO feature in the power spectrum and estimate the BAO scale precision expected for different surveys by conducting a simple BAO fitting.

There have been a few studies that investigated the feasibility of the weak lensing tomography for detecting the matter BAO scale, which can be compared to our predictions for PWL surveys. \citet{Simpson2006} tested an $\ell$-dependent redshift binning of source galaxies to make the lensing kernel oscillate with the line-of-sight contribution of the BAO feature to reduce the projection effect; using a Fisher matrix approach, they predicted a $2-\sigma$ detection of the BAO feature from an LSST-like survey but with the number density and the redshift precision that are better than our default PWL-Stage IV. \citet{Grassi14} studied the possibility of detecting BAO from 3D photometric weak lensing \citep{Heavens03} using the Fisher matrix analysis, by measuring the significance of the power spectrum amplitude constraints in the locations of the segmented wiggles. Their approach is different from our approach since we are directly constraining the shift of the BAO scale. They found that the Eulid survey would be able to detect the amplitude change of the power spectrum with high significance for the wiggle feature at $k<0.1\hMpc$, while the detection quickly would become challenging when scales over $k<0.15-0.2\hMpc$ are included. The latter aspect of their result appears broadly consistent with our finding that the PWL-Stage IV cannot constrain the BAO scale with a meaningful precision.

We organize the rest of paper as follows. In \S~\ref{sec:methods}, we discuss the method of simulating shear power spectrum with tomographic bins and its covariance matrix and extracting the spatial power spectrum. In \S~\ref{sec:results}, we conduct a BAO fitting to the reconstructed spacial power spectra and report the resulting BAO precisions for various surveys. We vary the condition of the KWL surveys to investigate the effect of the shape noise and the redshift errors. In \S~\ref{sec:con}, we conclude.

\section{Methodology}\label{sec:methods}
In this section, we first describe how we generate mock cosmic shear data, i.e. angular auto and cross power spectra $C_\ell^{ij}$ from $i, \ j$ tomographic bins, as well as its covariance matrix. Again, we abbreviate the photometry-based weak lensing as `PWL', and the spectroscopy-based~\footnote{While a KWL survey requires an accompanying photometric survey on the overlapping area, the main observational resource will be dominated by the spectroscopic side of the survey due to the low source number density.} kinematic method as `KWL'. We simulate $C_\ell^{ij}$ with and without BAO information, respectively. From the mock shear power spectrum, we extract the spatial matter power spectrum $P(k)$ using the singular value decomposition (SVD) method. We abbreviate the power spectrum with and without BAO signal as $P_{\text{wig}}$ and $P_{\text{now}}$, respectively. Finally, we conduct a BAO fitting to the extracted BAO signal from the ratio of $P_{\text{wig}}$ over $P_{\text{now}}$.

\subsection{Simulating cosmic shear power spectrum data}
\begin{table*}
\begin{threeparttable}
\captionof{table}{Summary of parameters of default weak lensing surveys in our study. `KWL' stands for the kinematic weak lensing survey and `PWL' stands for the photometric weak lensing survey. The parameter setting is similar to Table $2$ in Huff13. The value of $n_{\text{gal}}$ below is the total angular number density of all tomographic bins.}
\begin{tabular}{ |l|c c c c c c c c| } 
 \hline
 \hline
 survey & area [$\text{deg}^2$] & $\sigma_{\epsilon}$ & $n_{\text{gal}}\, [\arcm]$ & $z_{\text{max}}$ & $z_{\text{mean}}$ & $z_{\text{med}}$ & $\sigma_z$ \\ 
 \hline
 KWL-Stage III & 5,000 & 0.021 & 1.1 & 1.3 & 0.62 & 0.59 & - \\ 
 KWL-Stage IV & 15,000 & 0.021 & 1.1 & 2.0 & 0.79 & 0.76 & - \\ 
 PWL-Stage III (DES) & 5,000 & 0.26 & 10 & 1.3 & 0.62 & 0.59 & 0.1(1+z) \\
 PWL-Stage IV (LSST) & 15,000 & 0.26 & 31 & 2.0 & 0.79 & 0.76 & 0.05(1+z) \\
 \hline
 \hline
\end{tabular}\label{tab:surveys}  
\begin{tablenotes}
\item[*]For the PWL-Stage III, we only consider photometric redshift up to $z=1.3$ while we allow the true galaxy source distribution $p(z_{\text{ph}}|z)$ to extend to $z=1.66$. For the PWL-Stage IV, we truncate the photometric and true redshift distribution both at $z=2$. We set $z_{\text{min}}=0.0$ in Stage III surveys, and $z_{\text{min}}=0.05$ in Stage IV surveys.
\end{tablenotes}
\end{threeparttable}
\end{table*}

Under the Limber approximation \citep{Limber54, Kaiser92},
the auto and cross shear power spectra from tomographic redshift bins $i, j$ are expressed as
\begin{align}
C^{ij}(\ell)=\frac{9H_0^4\Omega_m^2}{4c^4}\int_0^{\chi_h}d\chi \frac{g^i(\chi)g^j(\chi)}{a^2(\chi)}P_{\delta}\left(k=\frac{\ell}{\chi}, z(\chi)\right),\label{eq:Cijl}
\end{align}
where $\ell$ is the angular wavenumber, $\chi (z)$ is the comoving distance to redshift $z$, $\chi^h$ is the comoving horizon scale, $P_{\delta}$ is the underlying matter power spectrum at $k=\ell/\chi(z)$~\footnote{Although \cite{LA08} suggests $k=(\ell+1/2)/\chi(z)$ as a more exact Limber approximation, this refinement is not important for our purpose since we use the same approximation for both generating mock cosmic shear data and extracting the spacial power spectrum. } at redshift $z$, $a(\chi(z))$ is the scale factor at $z$, and $g^i(\chi)$ is the normalized lensing efficiency of $i$th tomographic bin
\begin{align}
g^i(\chi)=\frac{\int_{\chi}^{\chi_h}d\chi^{\prime}n^i(\chi^{\prime})
(1-\chi/\chi^{\prime})}{\int_{0}^{\chi_h}d\chi^{\prime}n^i(\chi^{\prime})},\label{eq:lenseff}
\end{align}
where $n^i$ is the (normalized) angular galaxy number density distribution of the $i$th redshift bin. We have assumed the flat Universe.

Since we are interested in testing the BAO information in the cosmic shear data rather than revisiting the well-studied $\Omega-\sigma_8$ information from the overall shape of the nonlinear shear power spectrum (e.g., Huff13), we bypass the halofit modeling of the nonlinear overall shape of $P_{\delta}$. Instead we approximate the underlying matter power spectrum $P_{\delta}(\ell/\chi, \chi)$ as the time-dependent linear matter power spectrum, i.e. $P_{\delta}(\ell/\chi, \chi(z))=D^2_{+}(\chi(z)) P(\ell/\chi, 0)$ with the normalized growth function $D_{+}(\chi(z))$ while accounting for the nonlinear BAO damping at the median redshift when constructing $P(\ell/\chi)$. 

In detail, for the input matter power spectrum $P_{\delta}(k)$, we use the following Gaussian BAO damping model to account for the BAO damping caused by bulk flows during structure formation, i.e.
\begin{align}
P_{\delta}(k) = \big[P_{\text{lin}}(k) - P_{\text{sm}}(k) \big] \exp \big[ -k^2 \Sigma^2/2\big] + P_{\text{sm}}(k), \label{eq: Pwig_input}
\end{align}
where $P_{\text{lin}}(k)$ is the linear power spectrum calculated from CAMB~\footnote{http://camb.info} with cosmological parameters similar to the Planck15 \citep{Planck15} constraints: $\Omega_m=0.3075$, $\Omega_\Lambda=0.6925$, $\Omega_b h^2=0.0224$, $h=0.679$ and $\sigma_8=0.82$,
and $P_{\text{sm}}$ is the no-wiggle (without BAO) power spectrum calculated from the matter transfer function \citep{EH98}.
For a Stage III Dark Energy survey, we assume the nonlinear BAO damping scale $\Sigma=5.58\hMpc$~\footnote{We used Eq. 17 in \citet{Ding18} to calculate the nonlinear damping scale $\Sigma$.} defined based on $z\sim 0.65$; for Stage IV, we assume $\Sigma=4.75\hMpc$ at $z\sim1.0$ accounting for the different source redshift distributions between the two kinds of surveys. 

In order to reduce numerical error and better extract the small BAO information in the simulated data, we also input a dewiggled version of $P_{\delta,\rm now}(\ell/\chi, \chi) = P_{\text{sm}}$~\citep{EH98}, to simulate auto and cross $C_{\rm now}^{ij}(\ell)$ without the BAO feature, which will be compared to $C^{ij}(\ell)$.

We adopt 5--100 tomographic redshift bins ($\Nzbin$) for the KWL surveys. For example, $\Nzbin=30$ for the KWL-Stage III survey gives $\Delta z\sim 0.043$ for $z=0-1.3$, corresponding to $\Delta \chi\sim 88\hMpc$ at $z\sim 0.65$. 
$\Nzbin=100$ for the KWL-Stage IV survey gives $\Delta z\sim 0.0195$ for $z=0.05-2.0$, corresponding to $\Delta \chi\sim 50\hMpc$ at $z\sim 1$. 
Based on the convergence tests as we show later, we adopt results from $\Nzbin=100$ for KWL surveys and from $\Nzbin=30$ for PWL surveys as the default. 

The total number of auto and cross shear power spectra is $(\Nzbin+1)\Nzbin/2$ for a given angular scale $\ell$. Based on the assumed survey area, we set $\ell$ starting from $10$ for Stage III and $\ell$ from $4$ for Stage IV, respectively. Adopting $\Delta \ell=3$ and $\ell_{\text{max}}=2002$ for all survey stages, we have $665$ $\ell$ bins for Stage III and $667$ $\ell$ bins for Stage IV. At $z\sim 1$ and $k=0.1\ihMpc$, and therefore at $\ell \sim 230$, our choice of $\Delta \ell=3$ corresponds to $dk=0.0013\ihMpc$ at the given distance or $d\chi = 30\hMpc$ at the given $k$. 

\subsection{Covariance matrix of shear power spectrum}
Even without any other observational systematics, an observed shear power spectrum contains shape noise which comes from the intrinsic ellipticities of galaxies. If the shape noise is Gaussian-distributed, the effective shape noise in the two-point clustering in the $i$th tomographic bin is $\sigma_{\epsilon}^2/n^i$, where $\sigma_{\epsilon}$ is the rms intrinsic shear in each component as given in Table~\ref{tab:surveys}, and $n^i$ is the total number density (per steradian) of source galaxies in the $i$th tomographic bin. 
Hence, the observed shear power spectrum including the shape noise is 
\begin{align}
\hat{C}^{ij}(\ell) = C^{ij}(\ell) + \delta_{ij}\frac{\sigma_{\epsilon}^2}{n^i}.\label{eq:Clnoise}
\end{align}
For a given $\ell$, the data set consists of $\Nzbin(\Nzbin+1)/2$ auto and cross shear power spectra over all tomographic bins, i.e.
\begin{align}
\big\{ \hat{C}_{\ell}\big\} = (\hat{C}^{11}_{\ell},\, \hat{C}^{12}_{\ell},\, ...\,, \hat{C}^{1\Nzbin}_{\ell}, \, \hat{C}^{22}_{\ell}, ...\, , \hat{C}^{\Nzbin\Nzbin}_{\ell}),
\end{align}
where $\Nzbin$ is the total number of tomographic redshift bins. Using the Wick's theorem, we can derive the Gaussian covariance matrix of shear power spectrum~\citep[e.g.][]{HJ04, BH10}
\begin{align}
\mathbb{C}\big[\hat{C}^{ij}(\ell), \hat{C}^{pq}(\ell^{\prime})\big]=\frac{\delta_{\ell \ell^{\prime}}}{(2\ell+1)\Delta_{\ell} f_{\text{sky}}} \big[\hat{C}^{ip}(\ell)\hat{C}^{jq}(\ell)+\nonumber \\ 
\hat{C}^{iq}(\ell) \hat{C}^{jp}(\ell)\big],\label{eq:cov_shear}
\end{align}
where $f_{\text{sky}}$ is the surveyed fraction of the whole sky, which is given in Table~\ref{tab:surveys}.
The total dimension of the covariance matrix becomes $N_\ell\times\Nzbin\times(\Nzbin+1)/2$ by $N_\ell\times \Nzbin\times(\Nzbin+1)/2$, the inversion of which becomes numerically expensive and noisy with $N_\ell \simeq 668$ and $\Nzbin=100$. Utilizing the orthogonality of ${\delta_{\ell \ell^{\prime}}}$, we arrange the data and the covariance elements such that the covariance matrix is block-diagonal with $N_\ell$ blocks of each $\Nzbin\times(\Nzbin+1)/2$ by $\Nzbin\times(\Nzbin+1)/2$.

\subsection{Photometric weak lensing surveys}
In PWL surveys, source galaxies within each tomographic bin are selected based on their photometric redshifts and therefore the true underlying source redshift distribution $n^i(z)$ is subject to the photometric redshift errors. We model the probability distribution of photometric redshift $\zph$ given the true redshift $z$ as Gaussian with zero offset bias $z_{\rm bias}$ from $z$~\citep{Ma06}, i.e.
\begin{align}
p(z_{\text{ph}}|z)=\frac{1}{\sqrt{2\pi}\sigma_z(z)}\exp\bigg[-\frac{(z-z_{\text{ph}})^2}{2\sigma_z^2(z)}\bigg], \label{eq:p_zph}
\end{align}
where we set the rms $\sigma_z=0.1(1+z)$ for a Stage III survey and $\sigma_z=0.05(1+z)$ for a Stage IV survey, respectively. The true redshift distribution $n^i(z)$ of galaxies in the $i$th photometry tomographic bin ($z_{\text{ph}}^{i-1} \le \zph \le z_{\text{ph}}^i \big)$  is
\begin{align}
n^i(z) = \int_{z_{\text{ph}}^{i-1}}^{z_{\text{ph}}^i} dz_{\text{ph}} n(z) p(z_{\text{ph}}|z), \label{eq:ni_ph}
\end{align}
based on the Bayes' theorem. 
$n(z)\equiv \frac{d^2N}{dzd\Omega}$ in the integrand is the overall true galaxy distribution.

In this paper, we assume that PWL surveys mimic the DES (as Stage III) and the LSST (as Stage IV) and adopt the corresponding source galaxy number, the redshift distribution, and the survey area from Huff13, as listed in Table~\ref{tab:surveys}. In the left panels of Fig.~\ref{fig:nz}, we show $n(z)$ assumed for the PWL surveys with the thick black curves and the derived $n^i(z)$ with the dash-dotted ones. Due to the large redshift uncertainty, $n^i(z)$ extends over the wide range of redshift. $\Nzbin=30$ that we adopted for the PWL surveys based on the convergence test in \S~\ref{subsec:Nzbin} corresponds to $dz=0.043$ for the PWL-Stage III and $dz=0.065$ for the PWL-Stage IV, respectively. The lensing efficiency $g(\chi)$ for the PWL surveys is then derived by integrating $n^i(z)$ as in Eq.~\ref{eq:lenseff}.

\subsection{Kinematic weak lensing surveys}\label{subsec:KWL}
In KWL surveys, $p(\zph|z)$ is replaced by $p(\zspec|z)$ in Eq.~\ref{eq:ni_ph}. Due to the small redshift uncertainty from spectroscopy, we can assume binning the data into a series of narrow tomographic redshift bins (e.g. $\Delta z \sim 0.02$ for 100 tomographic bins) to fully utilize the available redshift accuracy/precision. For such a narrow redshift bin, we can assume the slowly varying $n^i(z)$ to be almost constant over $\zspec^{i-1} < z \le \zspec^{i}$. 
The true galaxy number density in the $i$th spectroscopic tomographic bin is then
\begin{align}
n^i(z) = n(z) \Pi_{\zspec^{i-1},\, \zspec^{i}}(z),
\end{align}
where $\Pi_{\zspec^{i-1},\,\zspec^{i}}(z)$ is the boxcar function with $1$ for $\zspec^{i-1} < z \le \zspec^{i}$ and $0$ otherwise. 
The lensing efficiency in Eq.~\ref{eq:lenseff} can then be reduced to

\begin{align}
g_{\rm KW}^i(\chi)&= 
\frac{\int_{\chi}^{\chi_h}d\chi^{\prime}n^i(\chi^{\prime})(1-\chi/\chi^{\prime})}{\int_{0}^{\chi_h}d\chi^{\prime}n^i(\chi^{\prime})}&
\nonumber\\
&= \frac{\bar{n}^i\int_{\chi}^{\chi_h}d\chi^{\prime}
(1-\chi/\chi^{\prime})\Pi_{\chi^{i-1},\chi^{i}}}{\bar{n}^i\int_{0}^{\chi_h}d\chi^{\prime}\Pi_{\chi^{i-1},\chi^{i}}} &\nonumber\\
&=\begin{bmatrix}
1-\chi \frac{\ln \chi_i - \ln \chi_{i-1}}{\chi_i - \chi_{i-1}}&,&
~\mbox{ if $\chi < \chi_{i-1}$}\\
\frac{\chi_i-\chi-\chi (\ln \chi_i - \ln \chi)}{\chi_i - \chi_{i-1}}&,& ~\mbox{  if $\chi_{i-1}\le \chi < \chi_{i}$}\\
0 &,& ~\mbox{  if $\chi_{i} \le \chi$}
\end{bmatrix},&
\end{align}
which will be used in
\begin{align}
C^{i\le j}(\ell)=\frac{9H_0^4\Omega_m^2}{4c^4}\int_0^{\chi_i}d\chi \frac{g_{\rm KW}^i(\chi)g_{\rm KW}^j(\chi)}{a^2(\chi)}P_{\delta}\left(k=\frac{\ell}{\chi}, z(\chi)\right).\label{eq:CijlKW}
\end{align}

Therefore the KWL efficiency truncates all contribution from underlying matter distribution beyond $z_i$ and for $k < \ell/\chi_i$ (or truncate $\ell >k\chi_i$ ). The neighboring redshift bins will truncate the information at different $z$ and $k$, unlike in the PWL surveys, although, as shown later in \S~\ref{subsec:optKWIV}, this little difference in the broad lensing kernels does not affect the BAO precision very much.

We adopt hypothetical spectroscopic survey parameters for the default KWL surveys that are very close to those derived in Huff13. A KWL-Stage III represents an experiment with current instruments, assuming a footprint of 5,000 square degrees. A KWL-Stage IV represents a future survey with an infrared spectrograph to extend to higher redshift than Stage III, which covers 15,000 square degrees, similar to the footprint of LSST.  The number density and the redshift distribution of source galaxies are modeled based on 
the COSMOS Mock Catalog \citep{Jouvel09} by applying criteria for robust shape measurement, disk galaxies, and line emissions within a wavelength range of the ground-based spectroscopy. The model distribution is subsequently sub-sampled at low redshift to match a more feasible target density of $1.1/{\rm arcmin}^2$.  Note that any KWL surveys will require an overlapping photometric survey, even though we focus on the gain from the spectroscopic data in this paper. More details can be found in \S 2.2 of Huff13. We also assume and test an optimistic version of KWL-Stage IV with $4.3\;\arcm$ in \S~\ref{subsec:optKWIV} as well as various choices of shape noise. 
 Therefore, while we reference the default survey conditions from Huff13, our results are general and straightforwardly translatable to other survey conditions that are different from Huff13 as we test a wide range of shape noise and target densities to identify the conditions for the BAO detection for both types of WL surveys in this paper.

\begin{figure*}
\includegraphics[width=0.45\linewidth]
{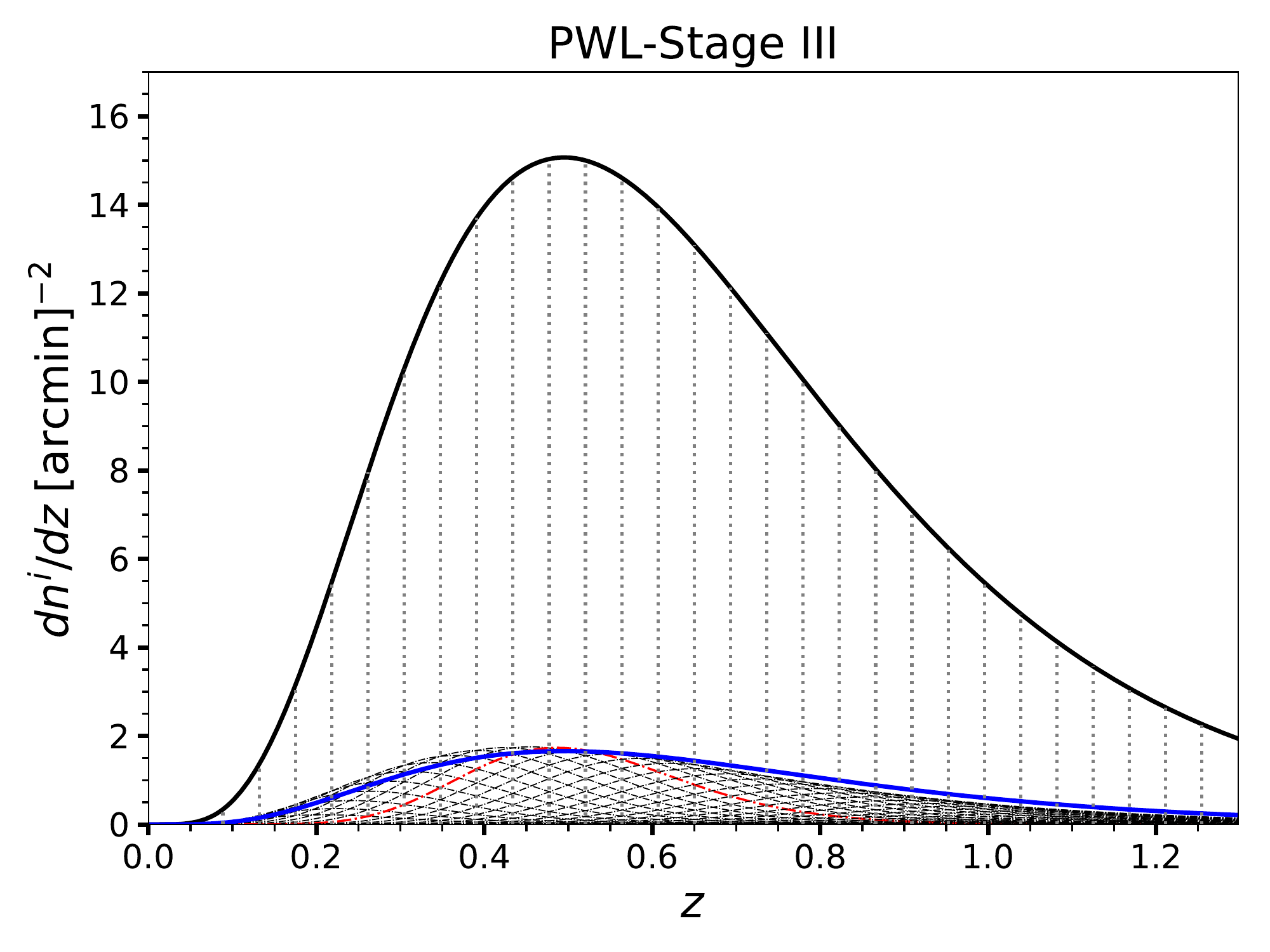}
\includegraphics[width=0.45\linewidth]{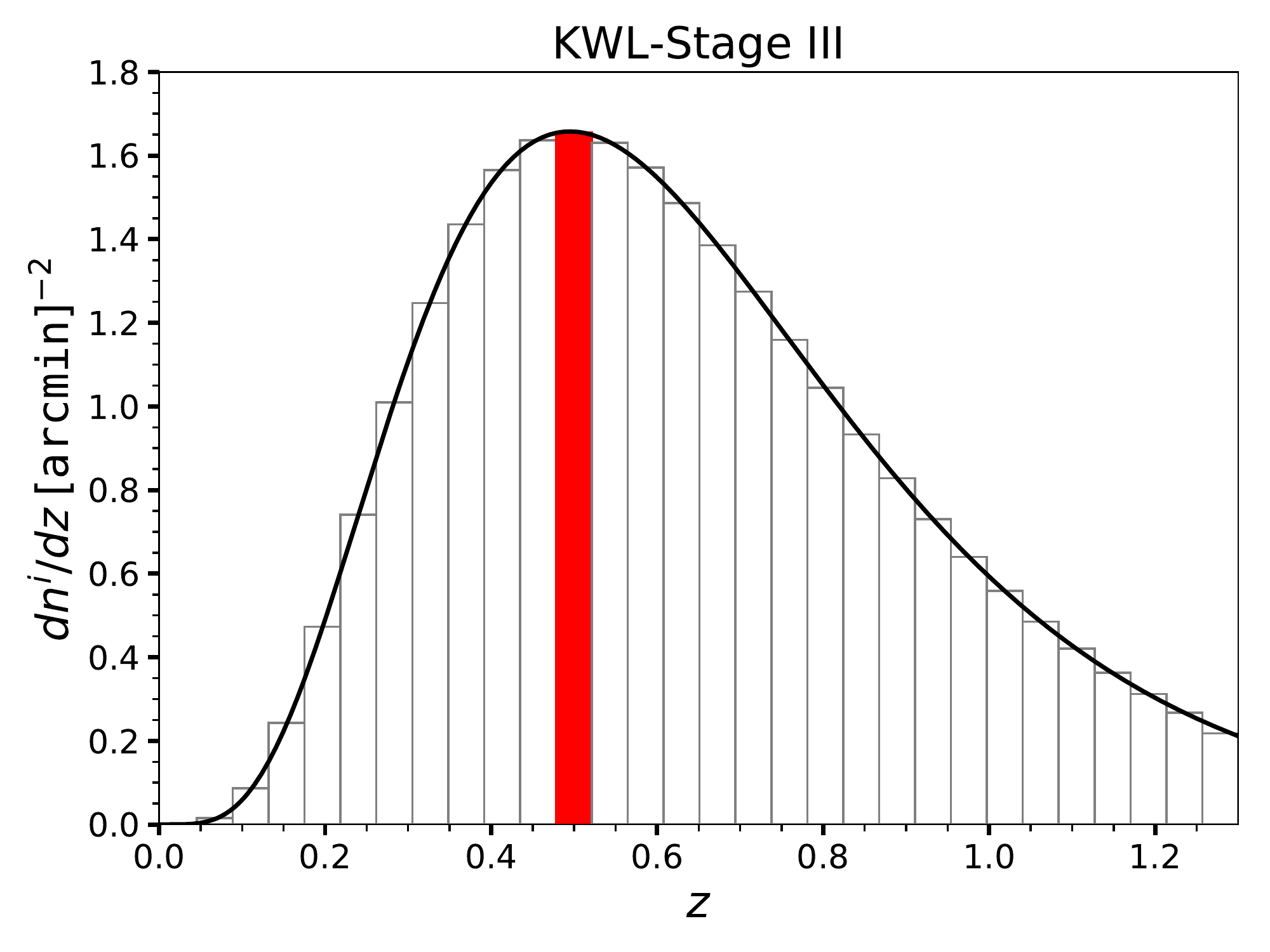}
\includegraphics[width=0.45\linewidth]{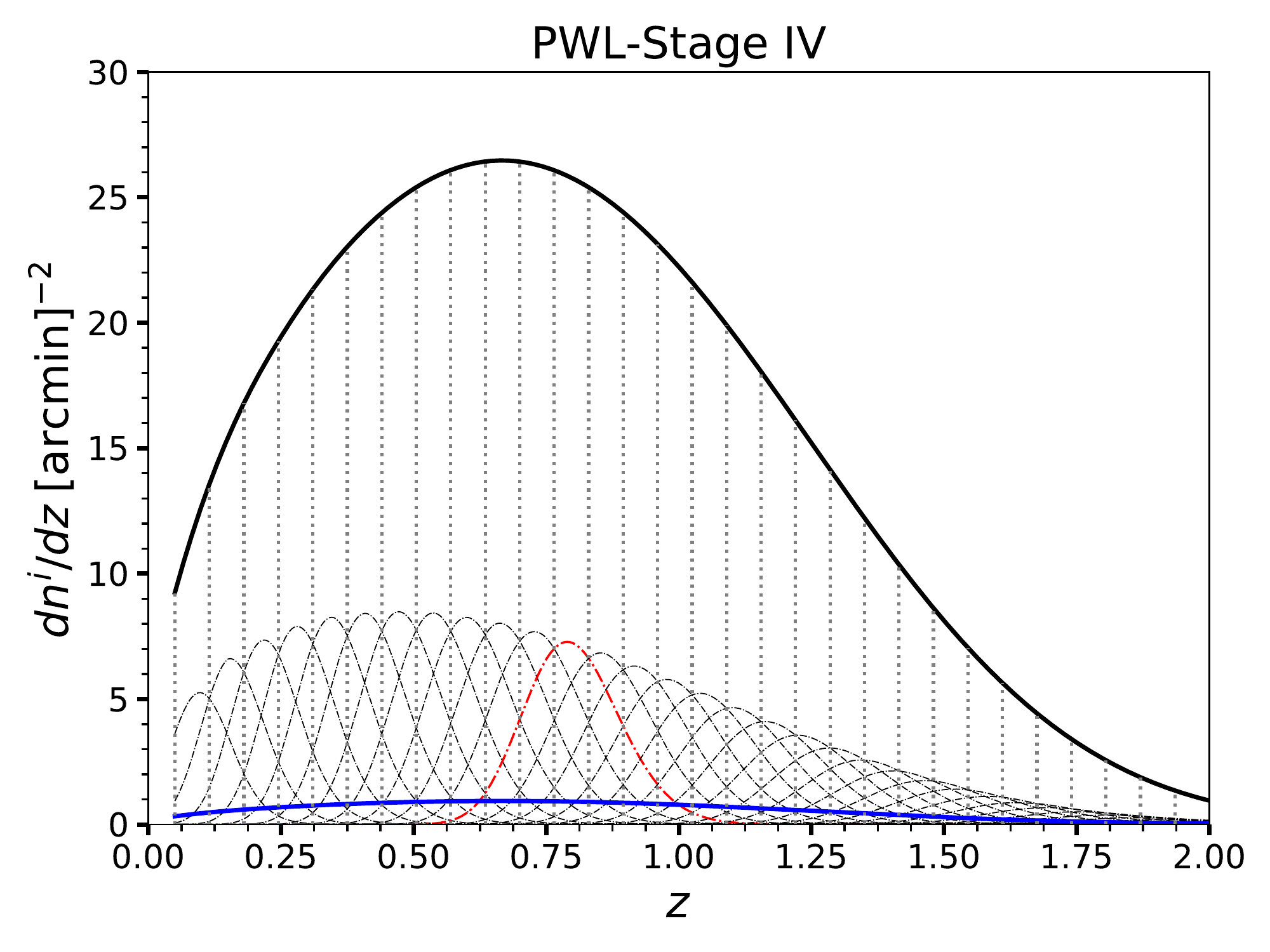}
\includegraphics[width=0.45\linewidth]{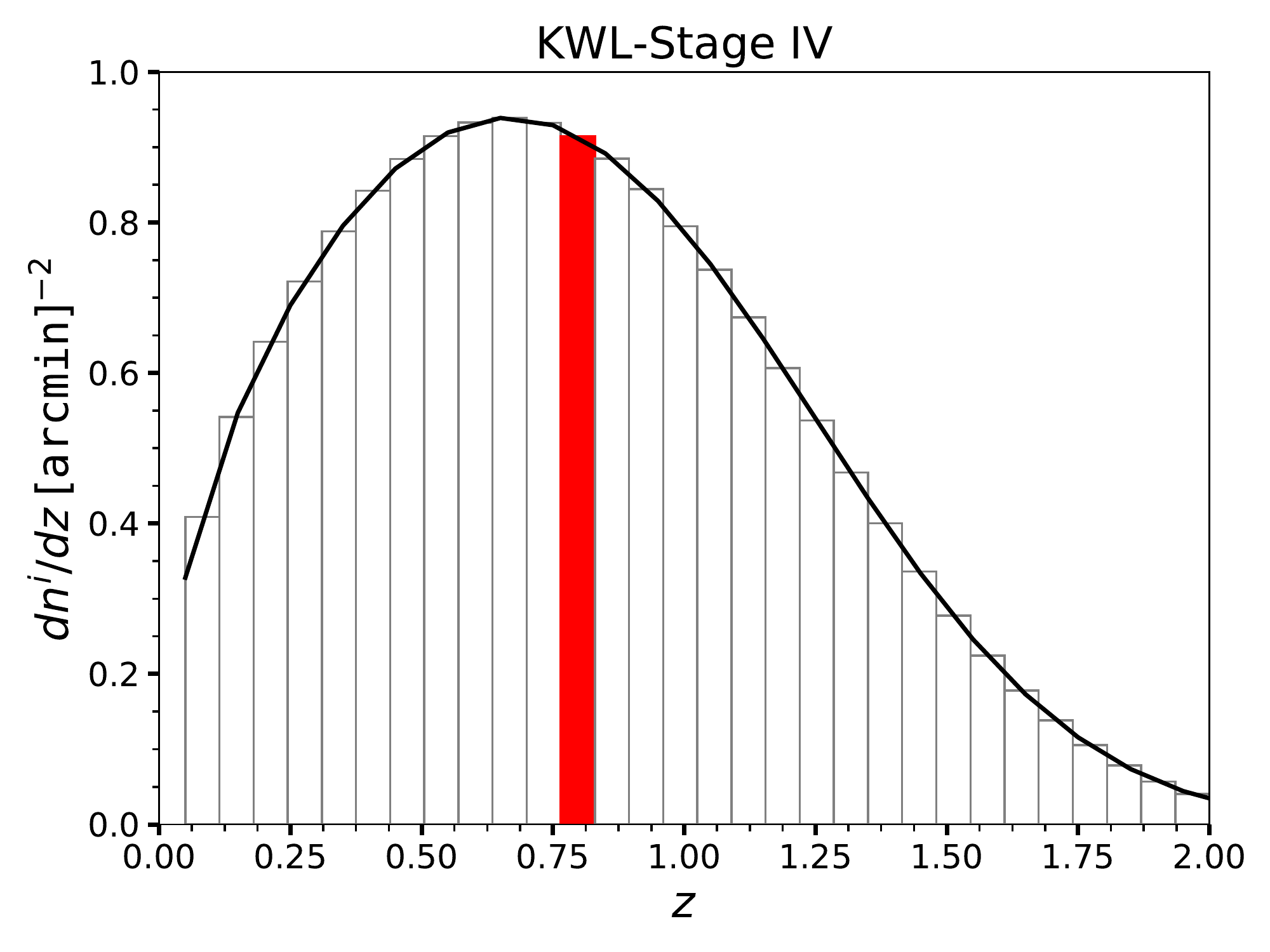}
\caption{The galaxy source distribution $n^i(z)$ in tomographic bins of PWL (left panels) and KWL surveys (right panels) based on Table \ref{tab:surveys}. Upper panels are for Stage III surveys and lower panels are for Stage IV surveys. To better visualize the difference between the two types of surveys, we assume $30$ evenly distributed tomographic bins (dotted lines in the left and histograms in the right) for both PWL and KWL surveys in this figure. The redshift range is $(0, \, 1.3)$ for Stage III and $(0.05,\, 2.0)$ for Stage IV, respectively. In the PWL surveys (left), each dash-dotted line represents $n^i(z)$ for each tomographic bin, while in the KWL surveys (right), it's represented by each histogram. We color $n^i(z)$ of one tomographic bin in red as an example. The thick black solid lines in all panels denote the overall true galaxy number density distributions. The overall distribution of the KWL surveys are also overlayed as blue solid lines in the left panels to emphasize the small number of sources required by the KWL surveys compared to the PWL surveys. While the number of source galaxies per area is greater for the KWL-Stage IV, its peak $dn/dz$ is lower than that of the KWL-Stage III due to its greater redshift range.
}\label{fig:nz}
\end{figure*}

\begin{figure*}
\includegraphics[width=0.45\linewidth]{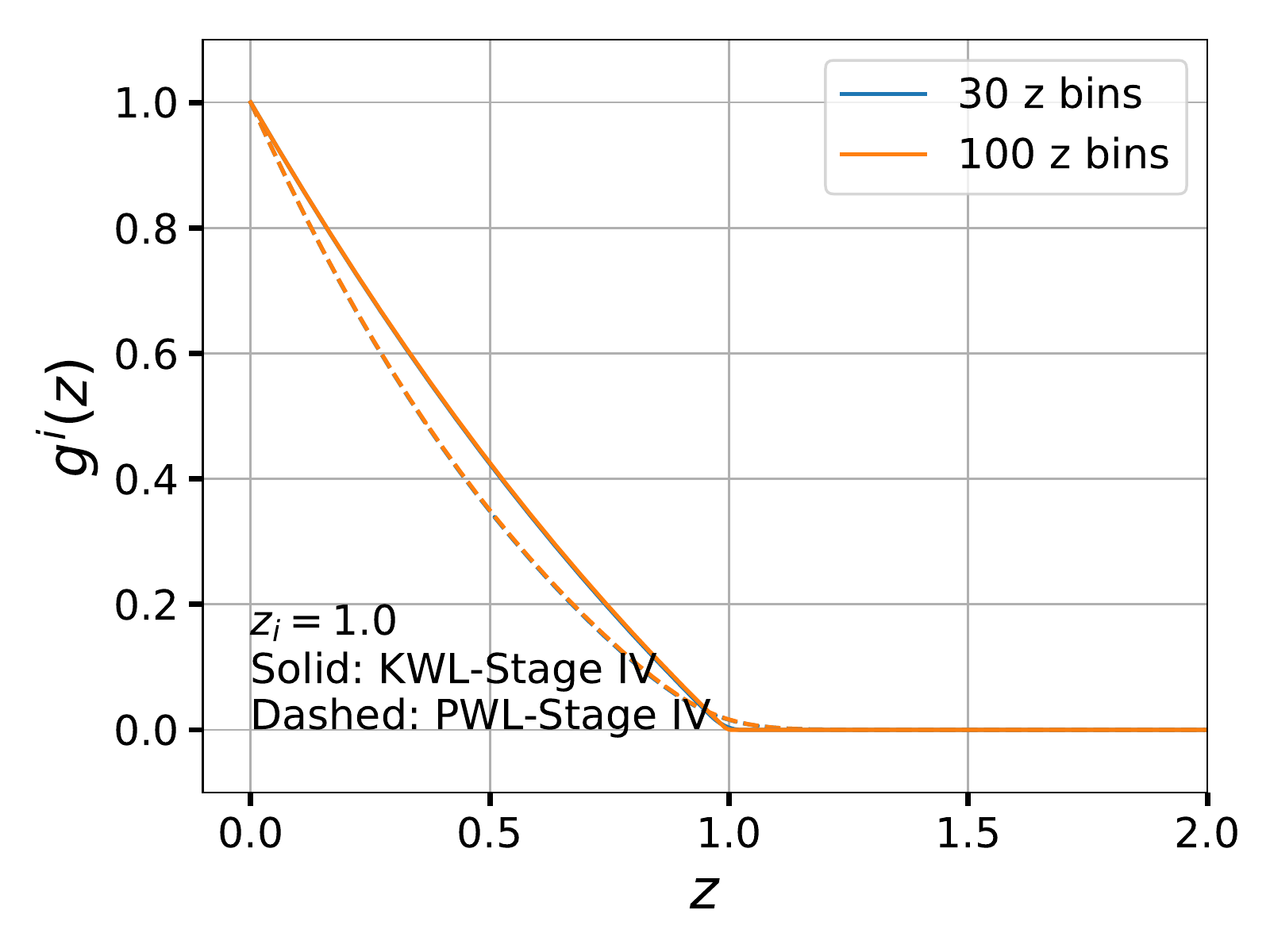}
\includegraphics[width=0.45\linewidth]{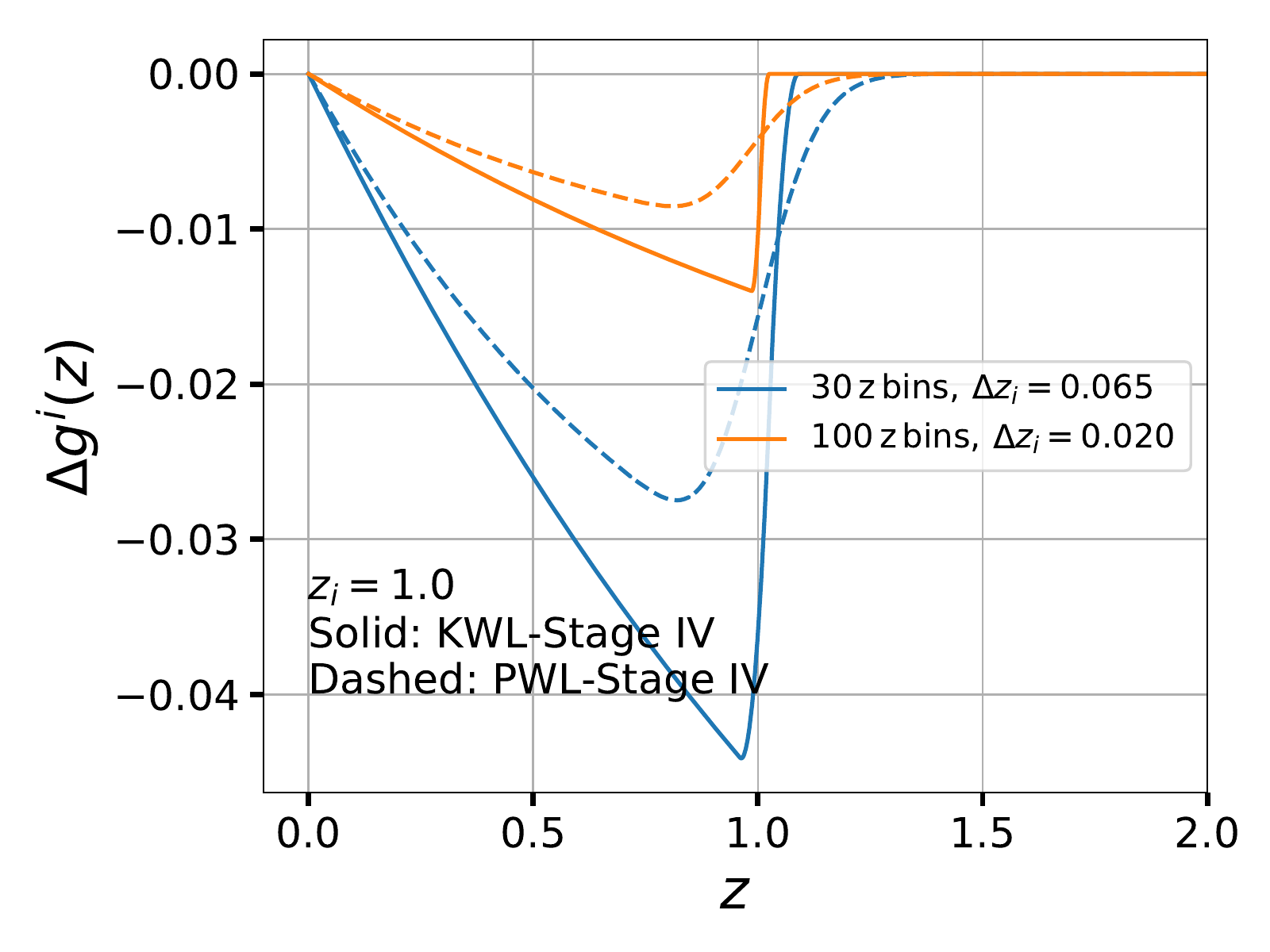}
\caption{The lensing efficiency kernel. \textit{Left:} the lensing efficiency kernel $g^i(z)$ of the $i$th tomographic bin centered at $z\sim 1$ for the  KWL-Stage IV (solid lines) and the PWL-Stage IV (dashed lines). One finds little difference between $\Nzbin=30$ (blue) and $\Nzbin=100$ (orange) for both surveys. \textit{Right:} the difference of lensing efficiency kernels ($\Delta g^{i+1}-\Delta g^{i}$) between neighboring tomographic bins around $z_i=1.0$. Given the same number of tomographic bins, the difference is slightly sharper for the KWL surveys. As the number of tomographic bins increases, the difference becomes smaller.
}\label{fig:leneff}
\end{figure*}

In the right panels of Fig.~\ref{fig:nz}, we show source galaxy distributions of the KWL (right panels) surveys in evenly distributed tomographic bins with $\Nzbin=30$ as an example. The black solid line in each panel shows the total true galaxy distribution. The same curves are also overlayed as blue solid lines in the left panels to emphasize the small number of sources required by the KWL surveys compared to the PWL surveys. We use $\Nzbin=30$ also for the KWL surveys to better visualize the difference between the two types of surveys. For the PWL surveys (left panels), each dash-dotted curve represents $n^i(z)$ for a given $i$th tomographic bin. Dotted vertical lines denote the boundaries of tomographic bins. For the KWL surveys, the tomographic bin is the same as $n^i(z)$. 
We see broad distributions of $n^i(z)$ in the PWL surveys compared to $n^i(z)$ of the KWL surveys. 
The resulting lensing efficiency kernels for the PWL-Stage IV and the KWL-Stage IV are shown in Fig.~\ref{fig:leneff}, where we also include results from $\Nzbin=100$. We find that despite the redshift accuracy, the lensing kernels of the KWL surveys are as broad as the PWL surveys, while the overall shapes are different. The right panel shows the difference between lensing kernels of neighboring tomographic bins for each survey. The difference appears slightly sharper for the KWL surveys and decreases with increasing $\Nzbin$.

\begin{figure*}
\includegraphics[width=0.95\linewidth]{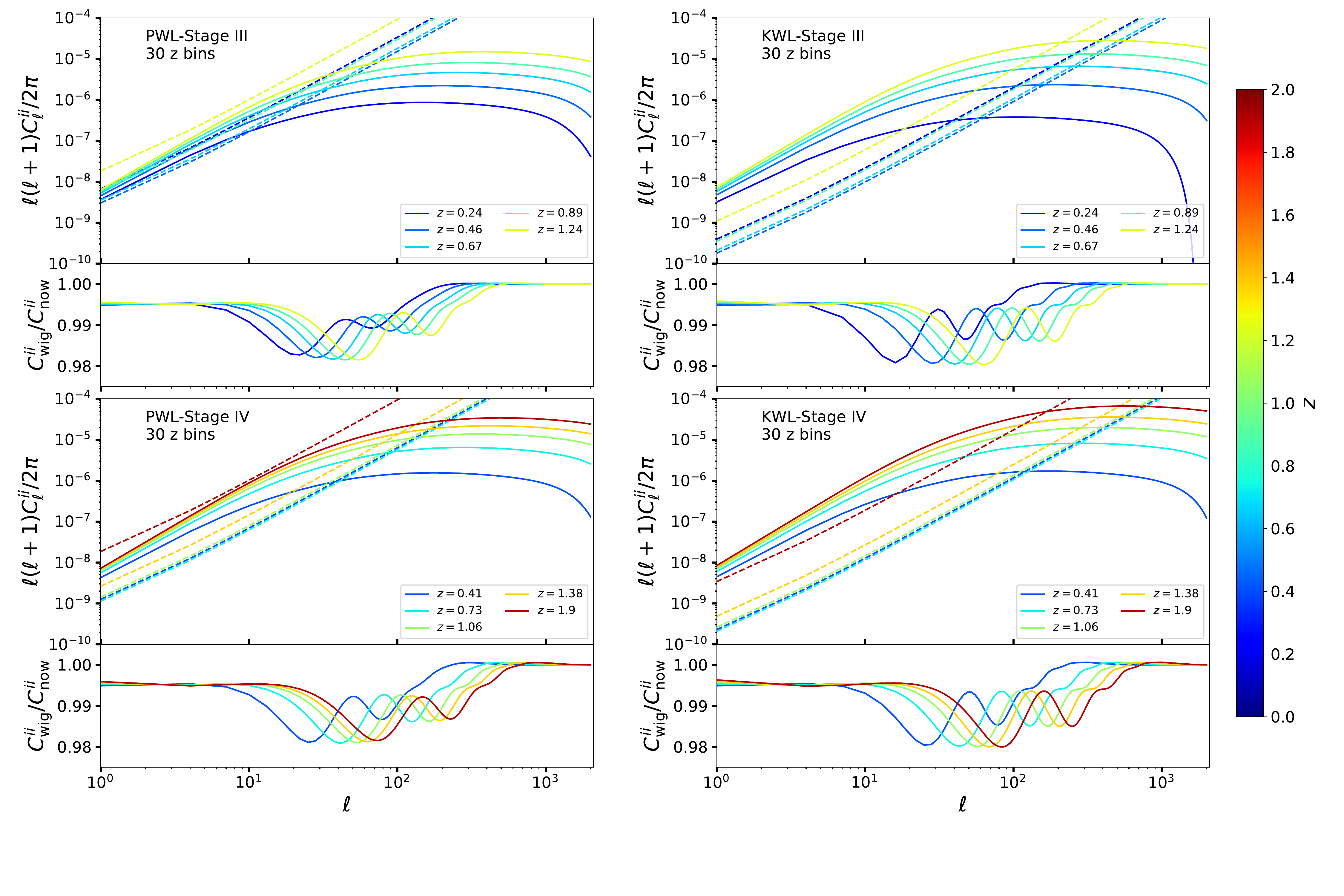}
\caption{The illustration of signal-to-noise ratio of shear power spectra for different surveys. We show $C^{ii}(\ell)$ (solid lines) and the corresponding shape noise $\sigma_{\epsilon}^2/n_i$ (dashed lines) at different redshift bins. Left panels show the PWL surveys and right panels show the KWL surveys. Top panels are for Stage III surveys and bottom panels for Stage IV surveys. Different colors denote different redshift bins. We assume $30$ tomographic bins for this figure. KWL surveys have higher signal-to-noise than the PWL surveys in terms of power spectrum amplitude. Increasing the number of tomographic bins linearly increases shape noise, without affecting the signal much. The lower part of each panel compares the BAO signal (wiggles) by taking the ratio of shear power spectra with and without containing the BAO signal. The amplitude of BAO wiggles appears larger in the KWL Stages, which implies that the KWL surveys potentially contain more BAO information in the signal.
}\label{fig:Ciil_shapenoise}
\end{figure*}

For each survey, we show auto shear power spectra $C^{ii}(\ell)$ (from Eq.~\ref{eq:Cijl}) and the corresponding effective shape noises $\sigma_{\epsilon}^2/n^i$ at different redshifts using $\Nzbin=30$ in Fig.~\ref{fig:Ciil_shapenoise}. 
The signal relative to the noise is much larger for the KWL surveys, reflecting the shape noise that is 17(5.4) times smaller than that of the PWL-Stage III(IV). The amplitude of the KWL auto power spectrum also tends to be higher than the PWL of the same tomographic bin, due to the greater area under the lensing kernels
 in Fig.~\ref{fig:leneff}. Therefore, another advantage of the KWL surveys for the purpose of the BAO extraction is their high signal-to-noise ratios of the overall amplitude.  
The second row of each panel in Fig.~\ref{fig:Ciil_shapenoise} shows $C^{ii}(\ell)$ divided by no-wiggle $C^{ii}(\ell)$ to single out the effect of the BAO feature in the shear power spectrum.
The figure implies that the BAO feature in the shear power spectrum is potentially stronger in the KWL surveys, again due to the shape of the KWL lensing kernel as will be discussed in \S~\ref{subsec:dependence}. This further increases the BAO signal to noise of the KWL surveys beyond the signal-to-noise ratio of the overall amplitude. 

\subsection{Extracting spacial power spectrum $P(k)$ from the simulated $C^{ij}_{\ell}$ data}\label{subsec:extPk}

We have an option to estimate the BAO information in the PWL and KWL surveys directly from the full set of  $(\Nzbin+1)\Nzbin/2$ auto and cross shear power spectra $C^{ij}_\ell$ given its $(\Nzbin+1)\Nzbin/2$ by $(\Nzbin+1)\Nzbin/2$ block-diagonal covariance matrix. For the data compression and for the visual clarification purpose, however, we choose to project the simulated $(\Nzbin+1)\Nzbin/2$ shear power spectra in the presence of the noise onto the underlying spacial power spectrum $P(k)$.

In order to extract $P(k)$ from the (mock) shear angular power spectrum $\hat{C}^{ij}(\ell)$ (i.e. which is already prepared using Eq.~\ref{eq:Cijl} and \ref{eq:Clnoise}) over a limited range of observable $\ell$, we approximate $\hat{C}^{ij}(\ell)$ as a discrete sum of weighted band power $P_\delta(k)$ instead of as the original integration (i.e. Eq.~\ref{eq:Cijl}):  
\begin{align}
{C}^{i\le j}(\ell) \simeq \frac{9H_0^4\Omega_m^2} {4c^4}\sum_{k_n>\ell/\chi_i}^{k_N} P_{\delta}({k}_{n}) \int_{1/(k_{n}+dk/2)}^{1/(k_{n}-dk/2)}d\bigg(\frac{\ell}{k}\bigg) \nonumber \\
\frac{g^i(\ell/k)g^j(\ell/k)}{a^2(\ell/k)} D_{+}^2(\ell/k) ,\label{eq:cal_cijl}
\end{align} 
where $k_{n}$ is the mean of the $n$-th $k$ bin, $N$ is the total number (i.e. the maximum) of $k$ bins, $z_n$ is the redshift of $\chi=\ell/k_n$, and $P_{\delta}({k}_{n})$ is the band power at $z=0$. 
We again approximate $\chi=\ell/k$ with the Limber approximation assuming the flat Universe. We assumed that the actual power $P_\delta(k)$ is smoothly varying within each $k$ bin and therefore can be pulled out of the integration. In order to have this assumption valid as well as to catch the BAO oscillation with the band powers, we use total $66$ output $k$ bins with the middle 64 bins logarithmically spaced in $k$ range $[0.01, \ 1.0]$ $\ihMpc$. The first and last bin is set as $10^{-4}<k<0.01$ $\ihMpc$ and $1.0<k<2.4343$ $\ihMpc$, respectively. which gives $dk\approx 0.0075\ihMpc$ at $k\sim 0.1\ihMpc$ and $dk\approx 0.015\ihMpc$ at $k\sim 0.2\ihMpc$.

Written in the matrix multiplication format,
\begin{align}
{C}^{ij}(\ell) = G^{ij}_{\ell k_n}*P_{\delta}(k_n),\label{eq:Cijmatrix}
\end{align}
where 
\begin{align}
G^{ij}_{\ell k_n} =\int_{1/(k_{n}+dk/2)}^{1/(k_{n}-dk/2)}\frac{g^i(\ell/k)g^j(\ell/k)}{a^2(\ell/k)} D_{+}^2(\ell/k) d\bigg(\frac{\ell}{k}\bigg).\label{eq:Gmatrix}
\end{align}

Taking the measurement error of $C^{ij}(l)$ as Gaussian distributed, we want to derive $P_{\delta}$ and its covariance by minimizing the $\chi^2$ of the likelihood function, i.e.
\begin{align}
\chi^2 = \left(\hat{C} - G*P_{\delta} \right)^T \mathbb{C}^{-1}\left(\hat{C} - G*P_{\delta} \right), \label{eq: chi2}
\end{align} 
where $\hat{C}$ denotes the observed/simulated (supposedly after shape noise subtraction) shear power spectrum, and the covariance matrix of $\hat{C}$, denoted as $\mathbb{C}$, is calculated from Eq.~\ref{eq:cov_shear}. The inverse of the covariance of $P_{\delta}$ that we need for the subsequent BAO scale fitting would be simply $G^{T}\mathbb{C}^{-1} G$, the second derivatives of $\chi^2$. We have an alternative to propagate this Fisher matrix to the BAO scale error using the Fisher matrix formalism, but again we choose to visualize $P_{\delta}$ and its errors and conduct a direct $\chi^2$ fitting to $P_{\delta}$.

In order to derive the maximum likelihood $P_{\delta}$ from ${C}^{ij}(\ell)$, we need to invert the non-square matrix $G$ in Eq. \ref{eq:Cijmatrix}. If we do not demand $G$ as a square matrix, $G$ is not invertible normally. Even if we set $G$ to be square, the limited range of data $\hat{C}(\ell)$ may not necessarily constrain all the band power $P_{\delta}$, resulting $G$ to be nearly singular. We could consider adding weak priors to such unconstrained band powers for better visualization of $P_{\delta}$ without affecting the final BAO scale constraint. In this paper, as one way of adding such weak priors, we take the singular value decomposition (SVD) approach and replace the small singular values with a minimum value of our choice if necessary. Compared to the typical way of adding a weak diagonal prior to the Fisher matrix of the $P_{\delta}$, the SVD with replacement corresponds to adding weak diagonal priors to poorly constrained eigen-vectors, i.e., combinations of $P_{\delta}$ in an eigen-vector space where the covariance and its inverse is indeed diagonal. We follow the method proposed by \citet{EZ01} (hereafter EZ01) and \citet{Pen03} to extract spatial matter power spectrum from angular power spectrum. We give a brief summery of the routine that we adopted. 

Note that in order to reduce the noise in the simulated data, we do not actually introduce random fluctuations due to shape noise or cosmic variance in generating $\hat{C}(\ell)$. The effect of the cosmic variance and the shape noise enters only in the covariance matrix $\mathbb{C}$. Following the methodology in EZ01, we rescale $P_{\delta}$  by a smooth function
\begin{align}
P_{\text{norm}}(k) = \frac{1.5\times 10^4 \trihMpc}{[1+(k/0.05\ihMpc)^2]^{0.65}}. 
\end{align} The resulting $P'_{\delta}= P_{\delta}/P_{\text{norm}}$ will allow equal fractional fluctuations on different $k$ scales, hence they receive similar weights when we apply the threshold for the singular values (SV) from the SVD that we perform later.

We can diagonalize the covariance matrix $\mathbb{C}$ by rotating it to its eigen-vector space. Since $\chi^2$ is a scalar, the value does not change by this rotation and therefore we conduct the matrix operation for $\chi^2$ in this eigenvector space: 
\begin{align}
\chi^2 = |\hat{C}'-G' P'_{\delta}|^2. \label{eq:chi2_modified}
\end{align}  
where $\hat{C}'=\mathbb{C}^{-1/2}  \hat{C}(\ell)$ and $G'=\mathbb{C}^{-1/2} GP_{\text{norm}}$ with $\mathbb{C}^{-1} = (\mathbb{C}^{-1/2})^{T}\mathbb{C}^{-1/2}$, where superscript $T$ denotes the transpose of matrix.
We therefore derive $\hat{C}'$ and $G'$ by transforming $\hat{C}$ and $G$ into the eigenvector space of $\mathbb{C}$ while scaling the eigenvector by the inverse square root of eigenvalues. In this way, we could apply SVD to derive $P^{\prime}_{\delta}$ that corresponds to the minimum residual in $|\hat{C}'-G' P'_{\delta}|$.

From SVD, we obtain
\begin{align}
 G' = U*W*V^T,
\end{align}
where $U$ and $V$ are column-orthogonal matrices, i.e., $U^T*U=I$ and $V^T*V=I$. Diagonal matrix $W$ stores all singular values (SV). See more details about SVD in \citet{Press92}.
Then the inverse of $G$ is 
\begin{align}
 G'^{-1} &= V*W^{-1}*U^T. \label{eq:G_inv}
\end{align}

As a result, we have
\begin{align}
P'_{\delta} = V * W^{-1} * U^T * C'. \label{eq:extract_P}
\end{align}

The inverse of the covariance matrix of $P'_{\delta}$ is then given by
\begin{align}
\text{Cov}^{-1}[P'_{\delta}] = G'^{T} G'.\label{eq:inv_covP}
\end{align}
\def\Nmodes{N_{\rm modes}}
In calculating  $W^{-1}$, therefore in calculating $P'_{\delta}$ and the covariance matrix, we check our results after replacing smaller SV with the minimum cutoff singular value $SV_c$ of our choice~\footnote{This is slightly different from EZ01, where the inverse of small singular values are replaced with zero when deriving $P'_{\delta}$. We find that replacing smaller SV with the minimum cutoff singular value $SV_c$ of our choice instead of zero results in less bias on the extracted $P'_{\delta}$ at the given number of SV modes. However, we find that both choices give the consistent BAO constraints once $P'_{\delta}$ is divided by $P'_{\delta, \rm {now}}$}. We vary choices of $SV_c$ and the corresponding number of unreplaced SV modes $\Nmodes$ and inspect the reconstructed BAO feature in $P'_{\delta}$ as well as the convergence in the BAO detection significance and the error constraint from the BAO fitting. We present results after such convergence is reached. Our main KWL BAO constraints quoted in the paper are using all SV modes without the $SV_c$ replacement (i.e., $\Nmodes=66$, or without any priors) while we explicitly note when we use a smaller number of unreplaced SV modes for the visual presentation purpose.

\begin{figure*}
\includegraphics[width=0.45\linewidth]{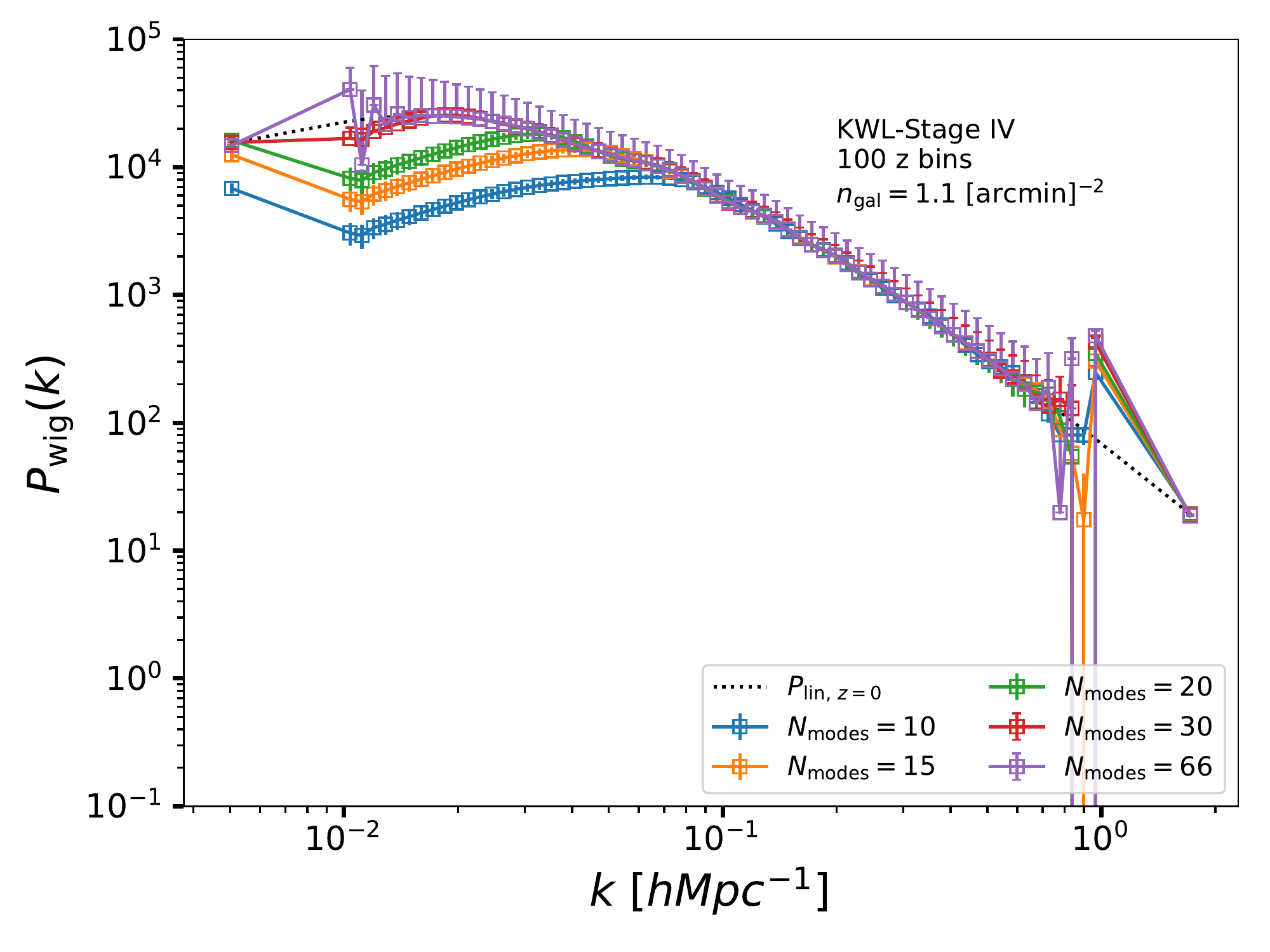}
\includegraphics[width=0.45\linewidth]{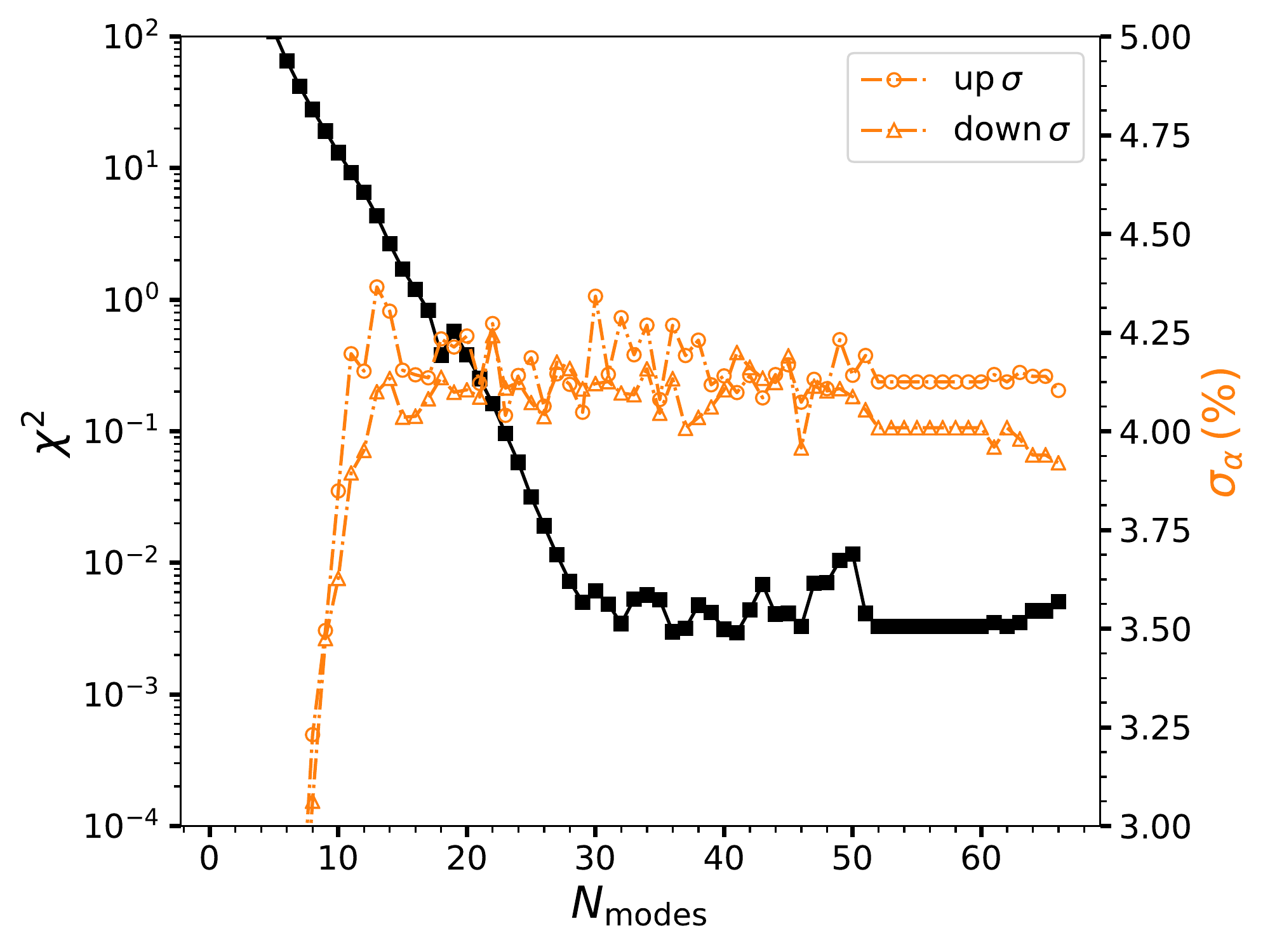}
\includegraphics[width=0.45\linewidth]{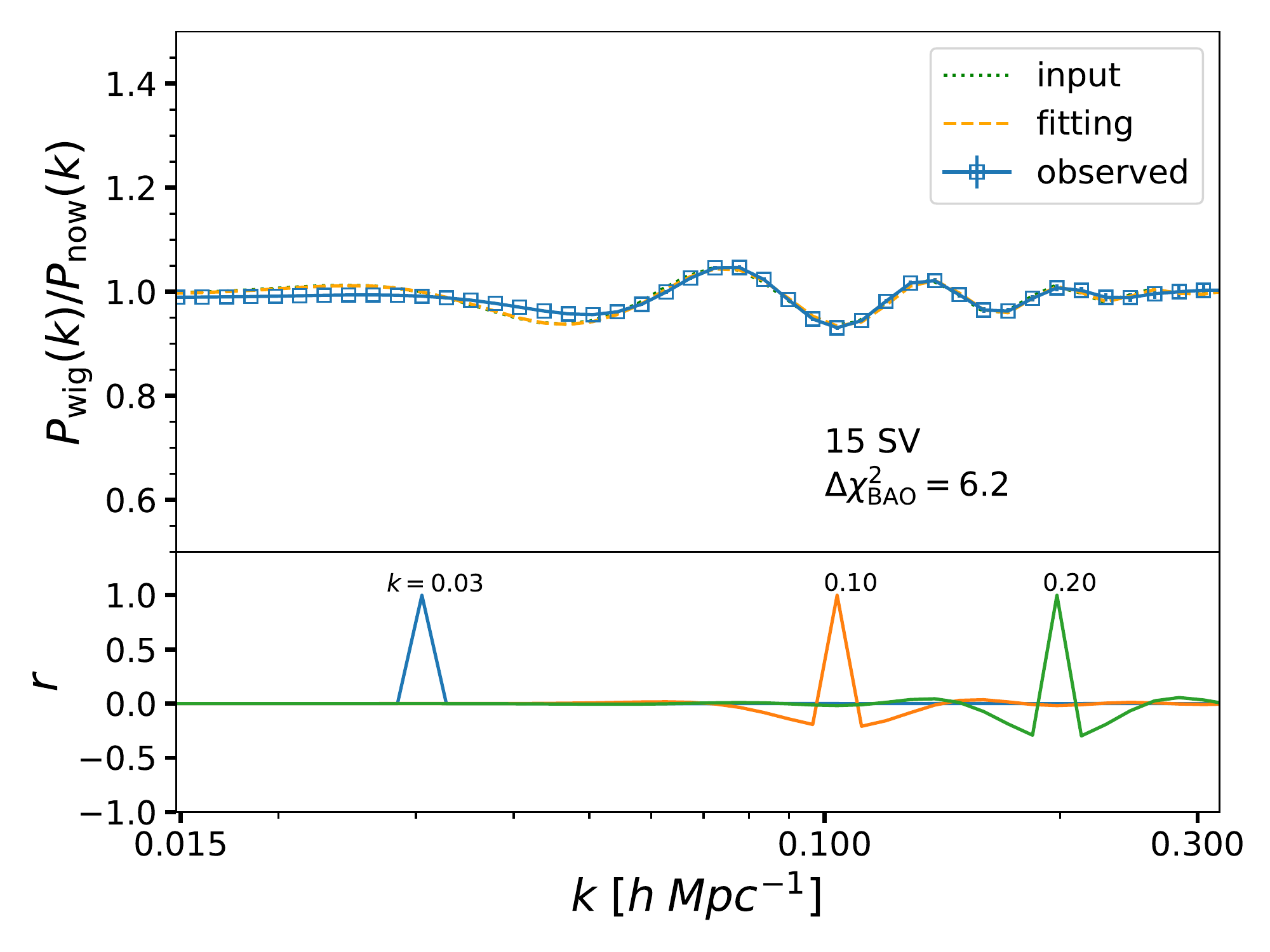}
\includegraphics[width=0.45\linewidth]{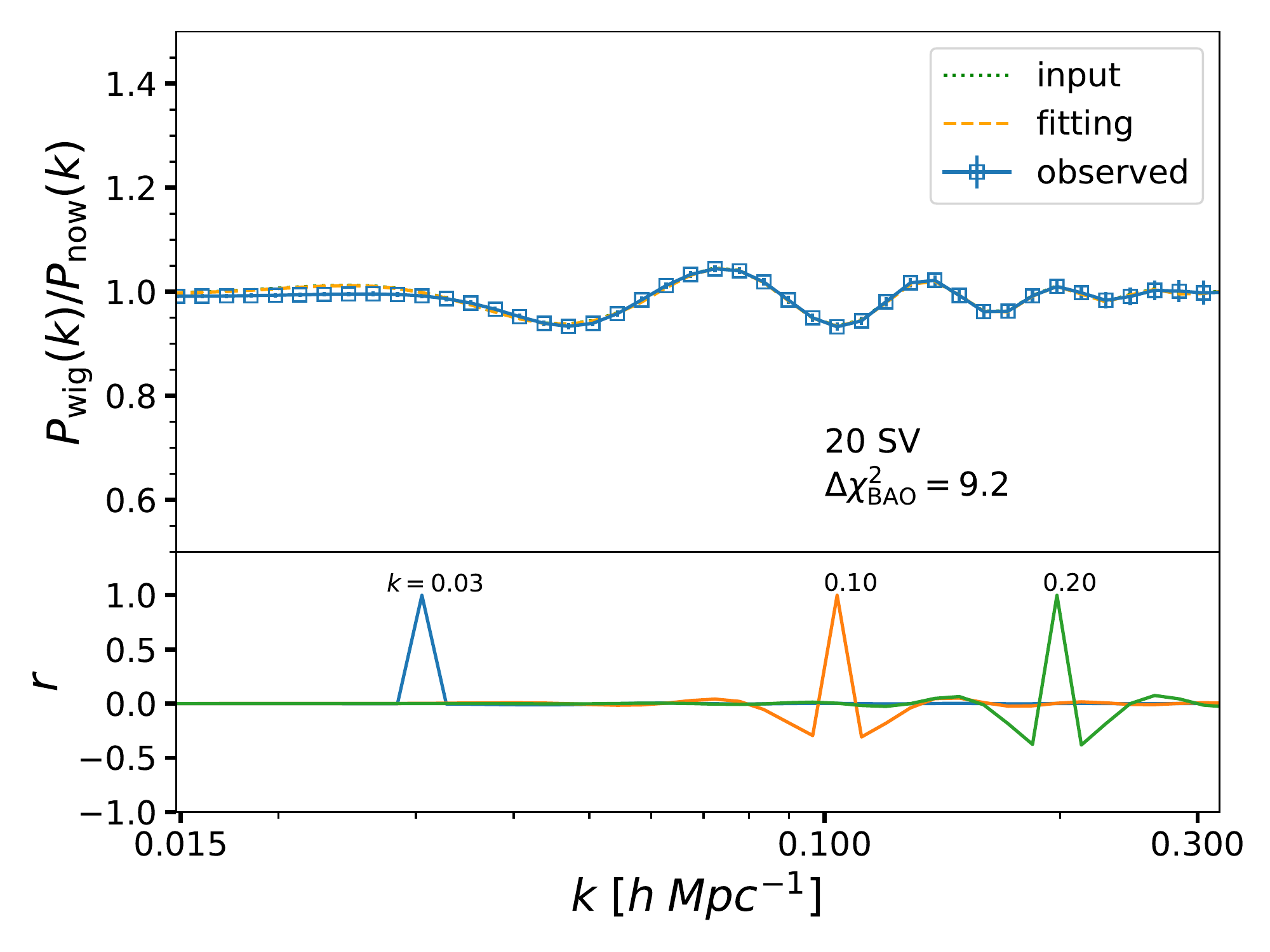}
\includegraphics[width=0.45\linewidth]{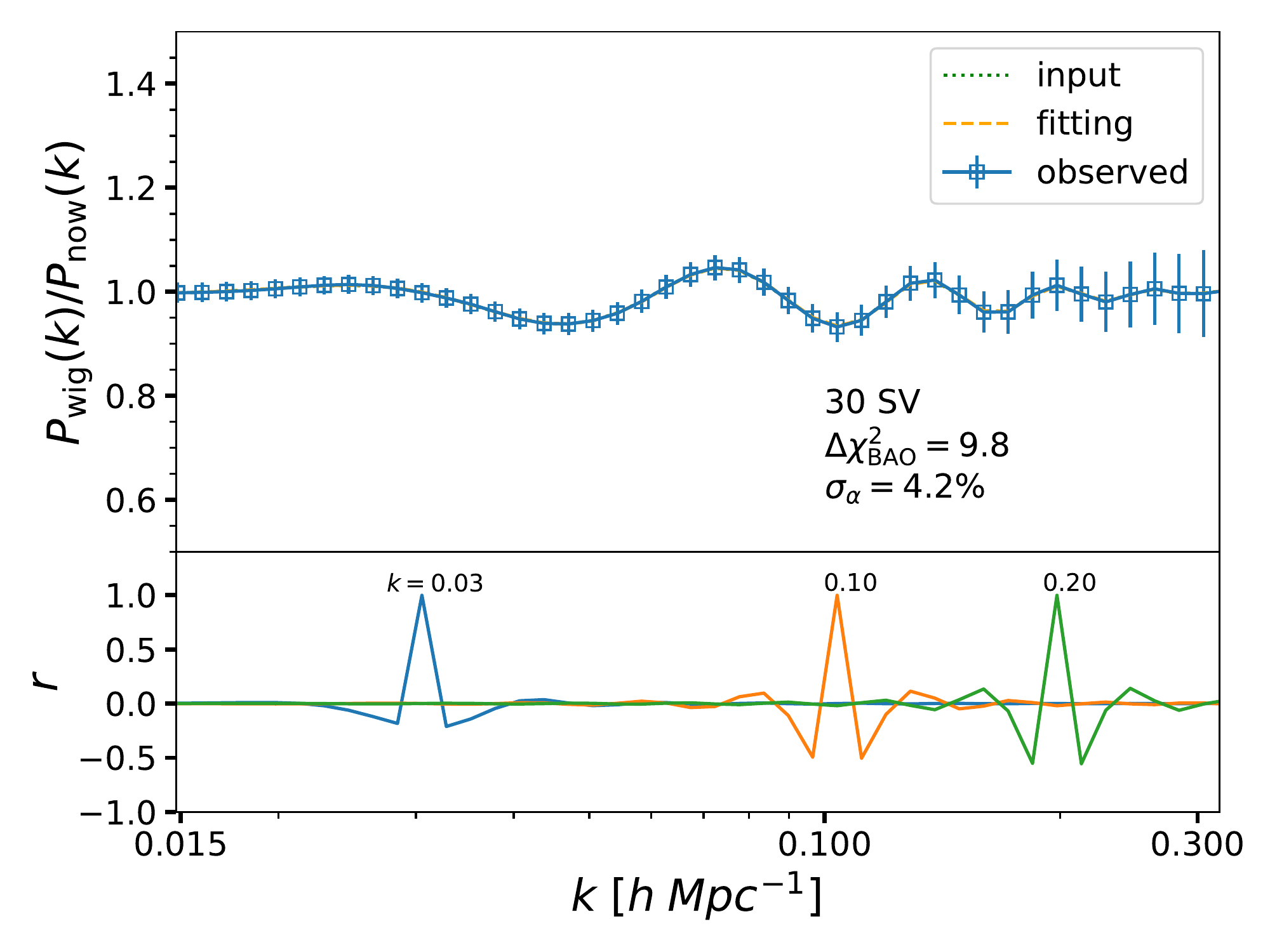}
\includegraphics[width=0.45\linewidth]{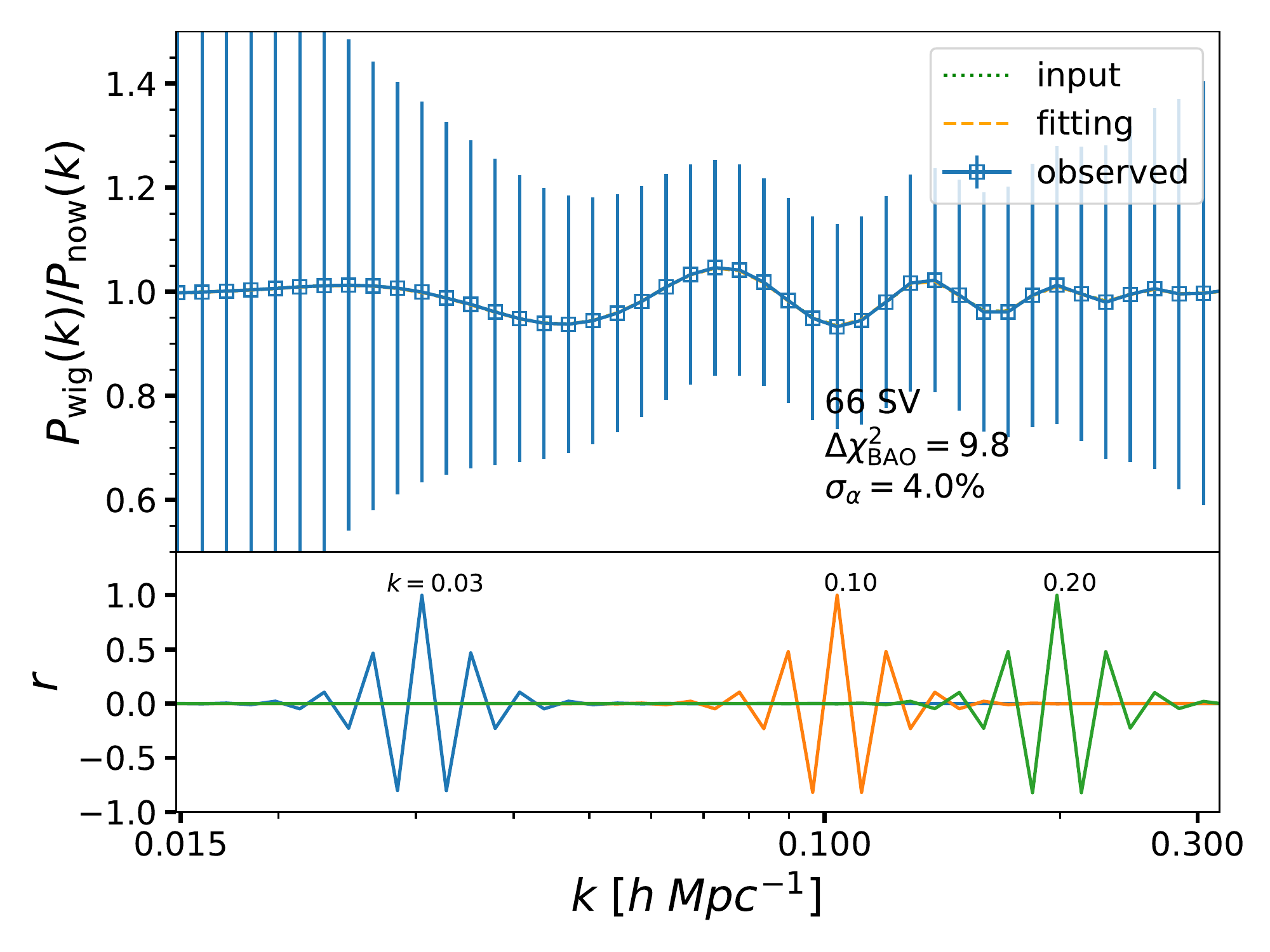}
\caption{Extracted power spectrum and covariance of the KWL-Stage IV as a function of the number of the unreplaced singular modes ($\Nmodes$) included in our calculation of Eq. \ref{eq:extract_P} and \ref{eq:inv_covP}. With $\Nzbin=100$. The top left shows the overall shape: including more modes without $SV_c$ replacement improves the small $k$ extraction of the overall shape. We only show the upper parts of the error bars for clarity. The middle and the bottom panels show the BAO feature as a function of $\Nmodes$; we also show the diagonal errors as well as the correlation coefficients $r=\text{Cov}_{ij}/\sqrt{\text{Cov}_{ii}\text{Cov}_{jj}}$ at $k_i=0.03$, 0.1, ad $0.2\ihMpc$ with different colors. In each upper panel, the solid line with empty squares denotes the extracted (observed) power spectrum, dotted and dashed lines denote input and fitted power spectrum, respectively. Again, increasing $\Nmodes$ better recovers the input BAO feature while it also increases the diagonal errors. Despite the increasing diagonal errors, different $k$ bins are more correlated with increasing $\Nmodes$ such that the BAO constraint reaches convergence once $\Nmodes > 30$ as shown in the top right panel. The upper and lower $\sigma_\alpha$ in the top right panel are derived from $\Delta \chi^2=\pm 1$ around the best fit $\alpha$ and we quote the average of these two values as $\sigma_\alpha$ in the main text and figures. For $\Nmodes=30$ and 66, we quote the BAO constraint $\sigal$ and $\Delta \chi^2$ (the detection significance squared) in the legends.
}\label{fig:KWIV_100z}
\end{figure*}

Fig.~\ref{fig:KWIV_100z} shows the effect of $\Nmodes$ in the extracted power spectrum as a function of $\Nmodes$ for KWL Stage IV, as an example. The top left panel  shows that the extracted power spectrum converges to the correct input broadband shape for $\Nmodes > 30$. The middle and bottom panels show, as the number of unreplaced modes increases, the reconstructed BAO feature converges to the input BAO feature. The top right panel shows that the BAO constraint and $\chi^2$ first decreases and then reaches convergence at $\Nmodes \sim 30$. Note that while the diagonal errors increase substantially beyond $\Nmodes = 30$, the off-diagonal covariance also changes such that the resulting error on the BAO scale does not change for $\Nmodes > 30$. In Fig.~\ref{fig:fit_Pwnw} we chose the minimum convergence $\Nmodes$ instead of the maximum $\Nmodes=66$ when visually presenting the extracted BAO feature to avoid the misleadingly large diagonal errors in some cases. 

\subsection{Fitting BAO wiggles}\label{subsec:baofitting}

In order to quantify the recovery of the BAO feature from KWL stage surveys, we constrain the BAO scales in the matter power spectrum reconstructed from the shear power spectrum. From the process described in Sec.~\ref{subsec:extPk}, we recover $P'_{\delta}$ (from the shear power spectrum with the BAO feature) and $P'_{\delta, \rm {now}}$ (i.e. from the shear power spectrum without the BAO feature). We fit the ratio of the two with a template of the BAO feature as
\begin{align}
\frac{P'_{\delta}(k')}{P'_{\delta,\text{now}}(k')} = A \bigg[ 1 +  \Big(\frac{P_{\text{lin}}(k)}{P_{\text{sm}}(k)} - 1\Big) \exp\big(- k^2 \Sigma^2/2\big) \bigg],\label{eq:fit_model}
\end{align}
with the presence of a free parameters $A$, which not only accounts for the constraint in the amplitude but also absorbs any effects that may rescale the amplitude of the recovered power spectrum, and $\alpha$, which relates the observed coordinate $k'$ and the template coordinate $k$ by $k=k'/\alpha$. 
Parameter $\alpha$ measures the shift in the BAO scale and therefore represents a constraint on comoving distance. A shift in $\alpha$ from unity would be originated from distortion in the comoving distance due to a wrong fiducial cosmology and therefore in principle such shift should rescale all the occasions of distances $\chi$ by $\alpha$ in Eq.~\ref{eq:CijlKW}~\footnote{To be more precise, a wrong cosmology will distort the distance in a redshift dependent manner.}. However, all other occasions of $\alpha$ will either cancel out or affect only the amplitude of the smooth kernel that multiplies to $P'_{\delta}$, not likely affecting the standard ruler test. We therefore only focus on the effect of $\alpha$ in $P'_{\delta}$ where $\alpha$ represents a characteristic shift in co-moving distance near the mean redshift of the survey. We also note that while our constraint on $\alpha$ is mainly from the transverse BAO feature due to the significant line of sight projection, as will be discussed later, it may not be straightforwardly interpreted as the BAO scale estimators from galaxy surveys such as the isotropic BAO scale $D_V$ or the angular diameter $D_A$.

In the fitting formula, we fix the BAO peak damping parameter $\Sigma$ as the input value. We omit nuisance parameters for additive or multiplicative nonlinearity in the shape of the power spectrum since our simulated shear power spectrum does not include nonlinearity except the BAO damping scale and also since such effect will be mostly cancelled out from the division by ${P'_{\delta,\text{now}}(k')}$.
In real observations, $P'_{\delta,\text{now}}(k')$ is not available. Our dividing by $P'_{\delta,\text{now}}(k')$ therefore would approximately correspond to a BAO-only fitting in real observations when the broad-band shape information is marginalized over with proper parameterization.
The advantage of the KWL survey on the broad-band shape information was extensively studied in Huff13.

We choose the fitting range in $[0.015,\, 0.3]$ $\ihMpc$. The covariance matrix for the limited $k'$ data is derived by inverting Eq.~(\ref{eq:inv_covP}) and then by taking the sub-covariance matrix. Such sub-covariance matrix is inverted again to find the best fit parameters and the corresponding errors through the $\chi^2$ analysis. Since sample variance is only included in the covariance matrix not in the mock data, the reduced $\chi^2$ would be negligible, much below unity, as long as the reconstructed power spectrum contains a reasonable BAO feature. We use the reduced $\chi^2$ as an indicator for the recovery of the BAO feature and the convergence of the fitting for various choices of $SV_c$, as shown in the top right panel of Fig.~\ref{fig:KWIV_100z}. 
We repeat a fitting with a no-BAO template, i.e., with $\Sigma = 100\hMpc$, 
and compare the $\chi^2$ curves of the BAO and no-BAO fittings to derive the detection level, $\sqrt{\Delta \chi^2}$. 
\\

\section{Results}\label{sec:results}

\subsection{The BAO information from the weak lensing surveys}
\begin{figure*}
\includegraphics[width=0.45\linewidth]
{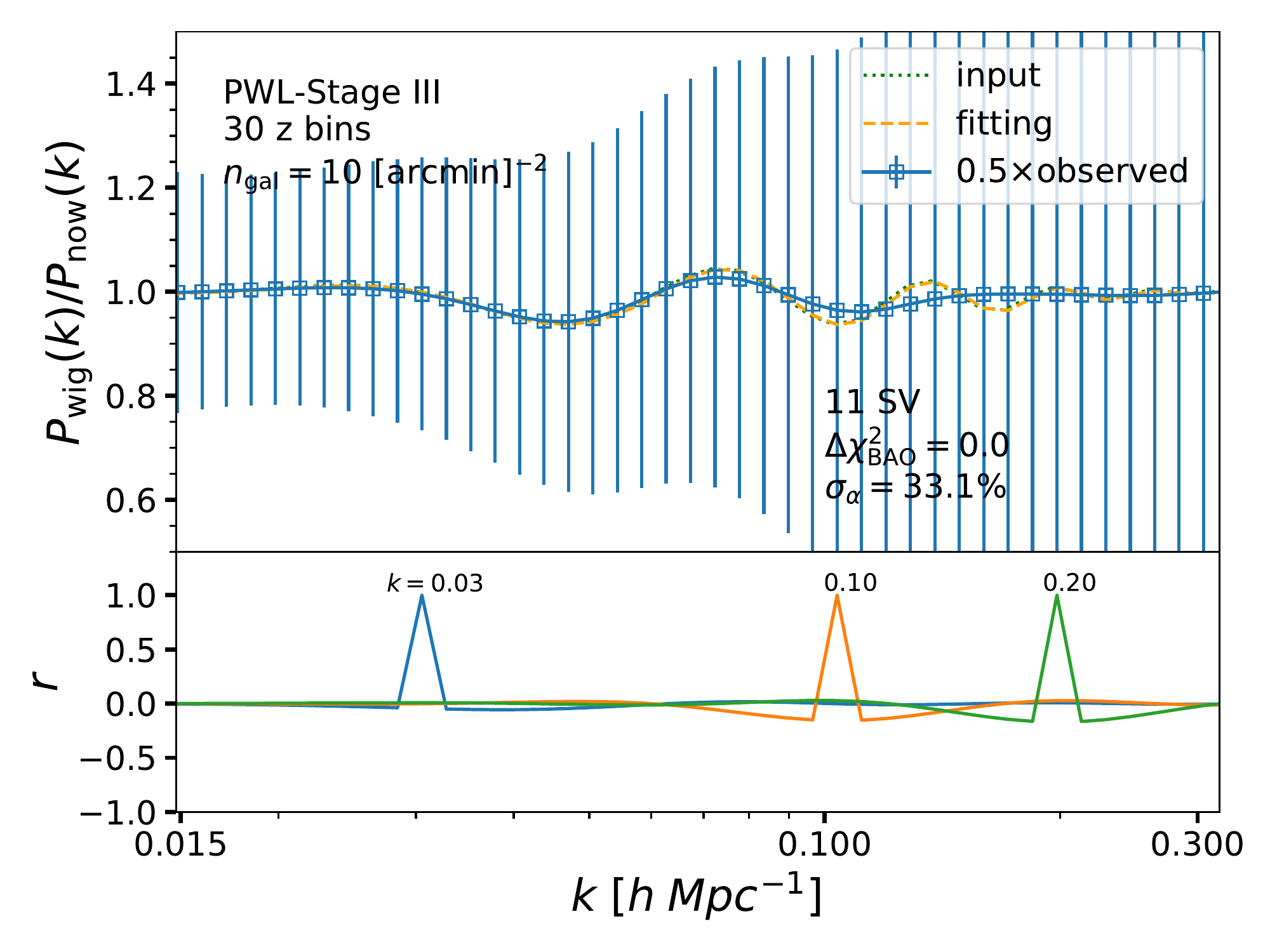}
\includegraphics[width=0.45\linewidth]{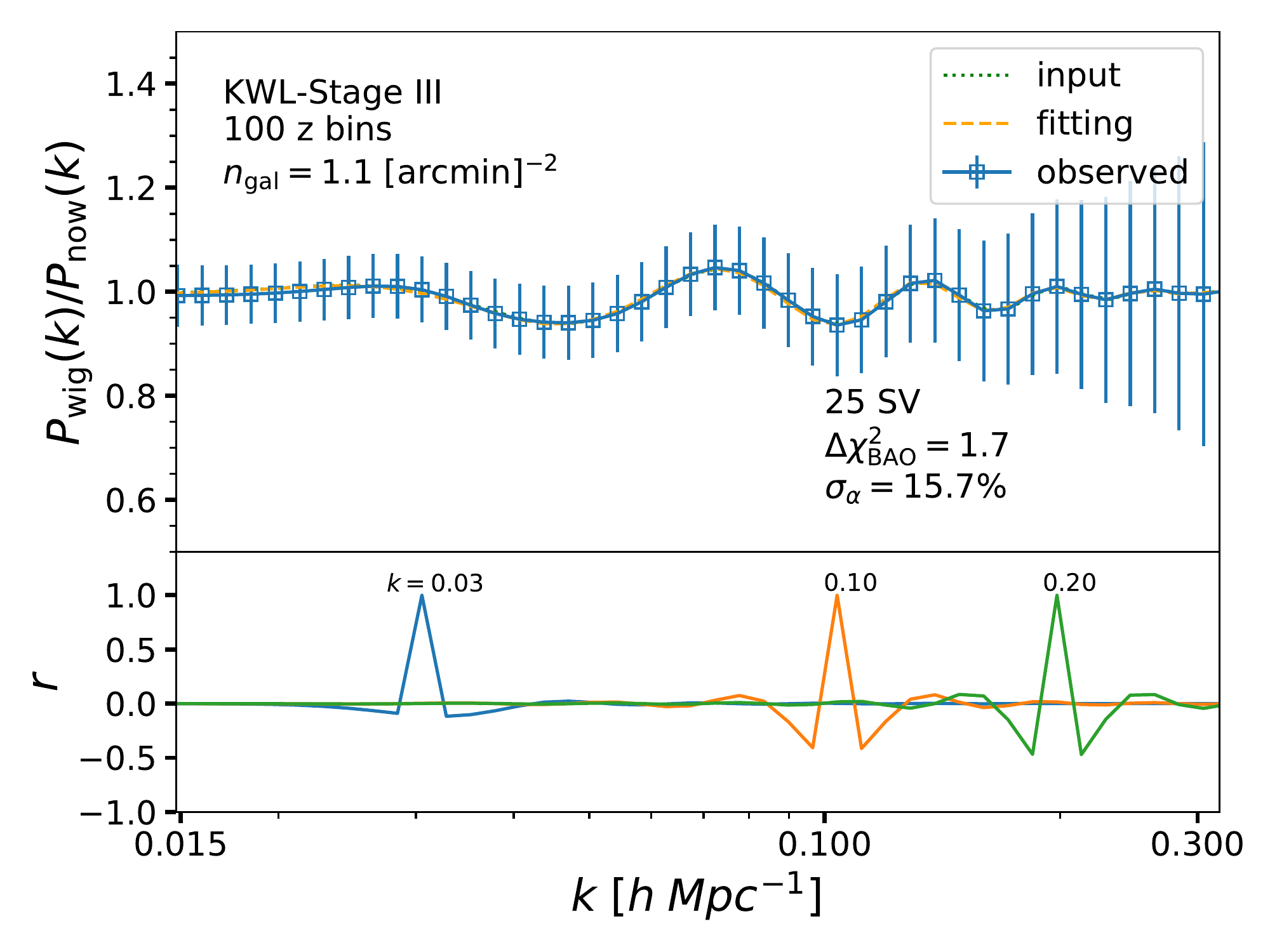}
\includegraphics[width=0.45\linewidth]
{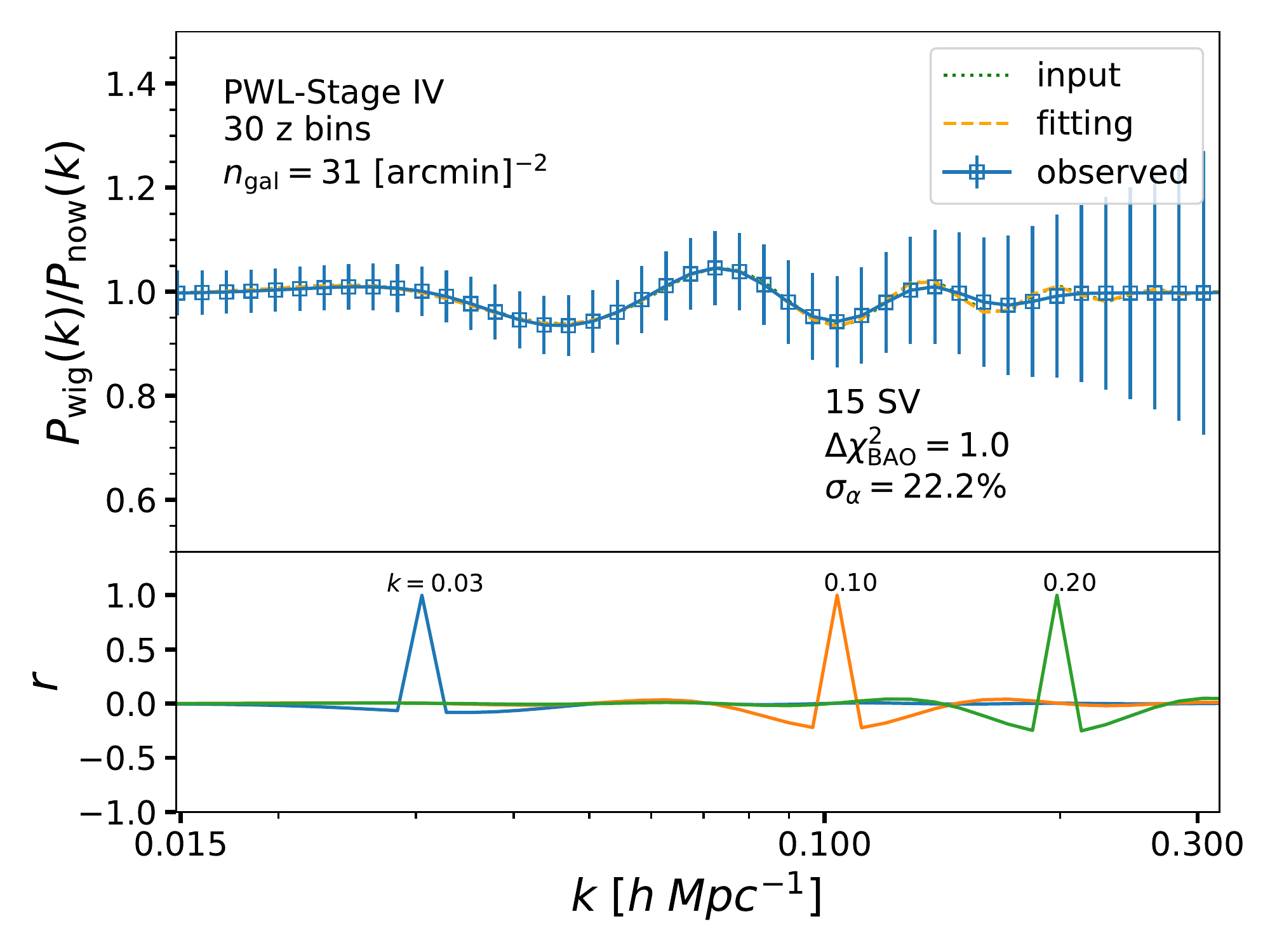}
\includegraphics[width=0.45\linewidth]{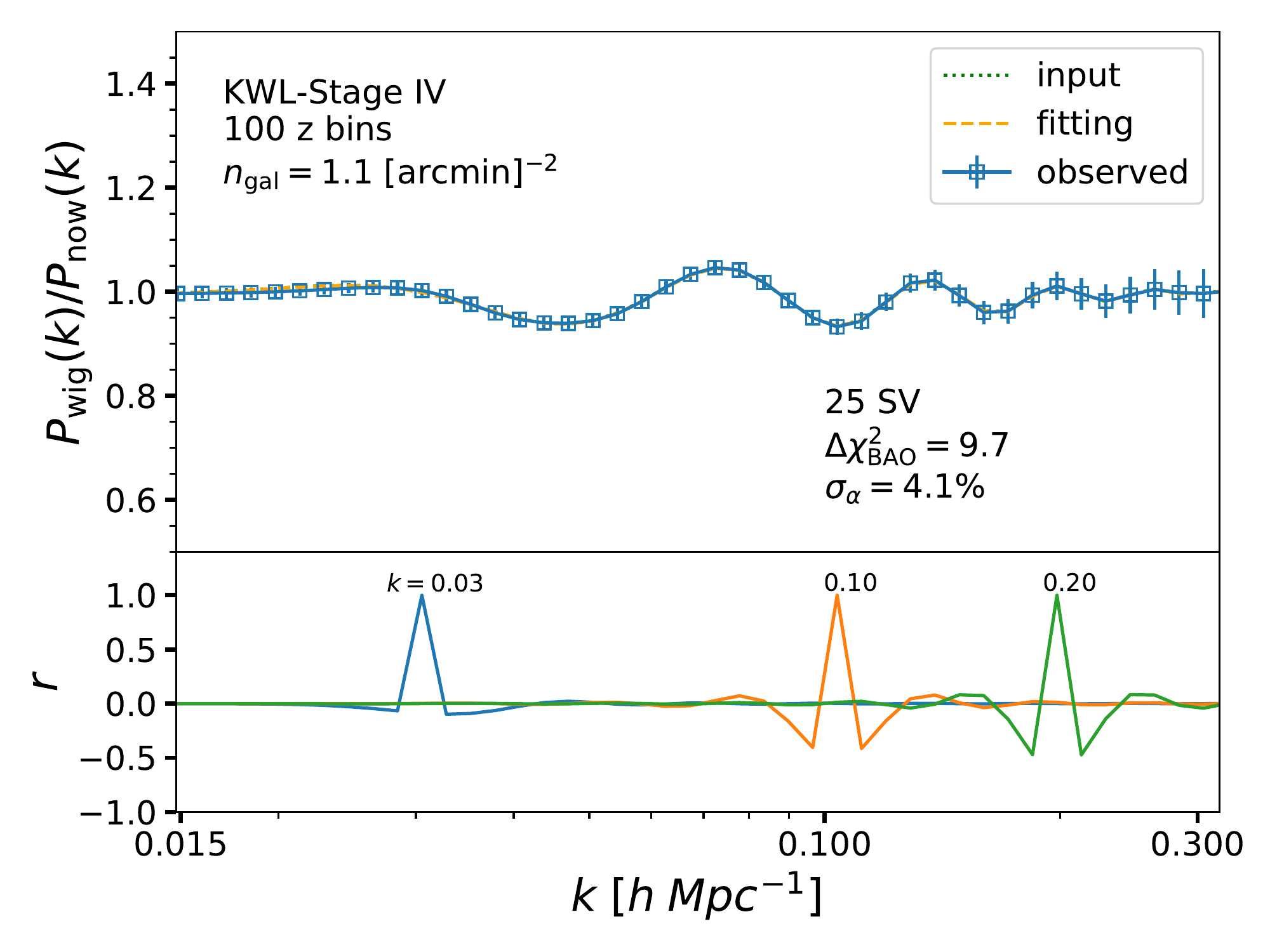}
\caption{The extracted matter power spectrum and its diagonal errors for PWL (left) and KWL (right) surveys from Table~\ref{tab:surveys}. Each panel represents one survey, in which, the upper part shows the band power ratio of $\Pwig$ over $\Pnow$. The black dotted line represents the input power spectra ratio, the square points with error bars denote the reconstructed power spectra ratio, and the dashed line denotes the best fitting; the lower part shows the correlation coefficients of the covariance matrix of the band power at $k=0.03$, $0.1$ and $0.2$ $\ihMpc$. 
We also note $\Nmodes$ (the number of unreplaced singular modes) used for extracting the power spectrum, along with the resulting precision on the BAO scale parameter $\alpha$ and the BAO detection significance squared $\Delta \chi^2_{\text{BAO}}$ from fitting. 
For the PWL-Stage III survey (upper left panel), we show the error bars of band power 2 times smaller than the true for clarity. 
The PWL surveys show the reconstructed BAO feature that are slightly more damped than the input BAO feature on large $k$ for this choice of $\Nmodes$; the difference decreases without affecting $\sigal$ as we increase $\Nmodes$.}\label{fig:fit_Pwnw}
\end{figure*}

In this section we present and compare the BAO feature in the power spectra reconstructed from the default PWL and KWL surveys specified in Table~\ref{tab:surveys}. In Fig.~\ref{fig:fit_Pwnw}, we show the ratio of the reconstructed spatial power spectrum $P_{\text{wig}}(k)$ over the reconstructed $P_{\text{now}}(k)$ and its diagonal errors (blue lines and error bars) and the best fit model (orange dashed line) in comparison to the the theoretical input (green dotted lines). The top panels are for Stage III and the bottom panels are for Stage IV. Since we do not introduce fluctuations in the simulated shear power spectrum itself, the reconstructed $P_{\text{wig}}/P_{\text{now}}(k)$ all show the BAO feature close to the input (i.e. the BAO damping scale $\Sigma=5.58\hMpc$ for Stage III and $\Sigma=4.75\hMpc$ for Stage IV) as long as a sufficient number of unreplaced SV modes are included.

\begin{figure}
\centering
\includegraphics[width=0.9\linewidth]{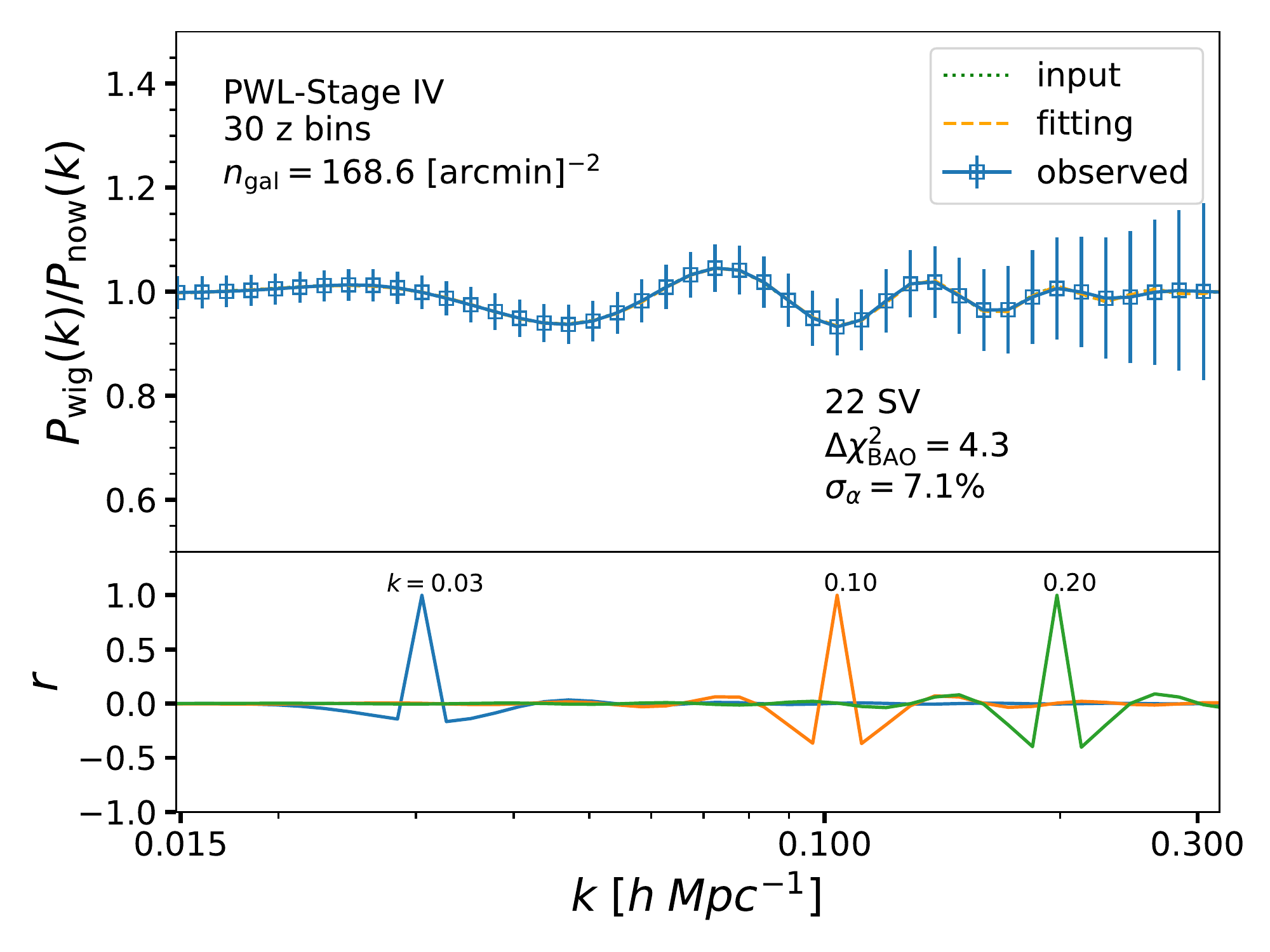}
\caption{The low noise PWL-Stage IV survey after decreasing the effective shape noise to match with that of the default KWL-Stage IV. Compared to the default PWL-Stage IV, $\sigma_{\alpha}$ decreases to $\sim 7\%$.}\label{fig:PW_lowsn}
\end{figure}

\begin{figure*}
\includegraphics[width=0.45\linewidth]
{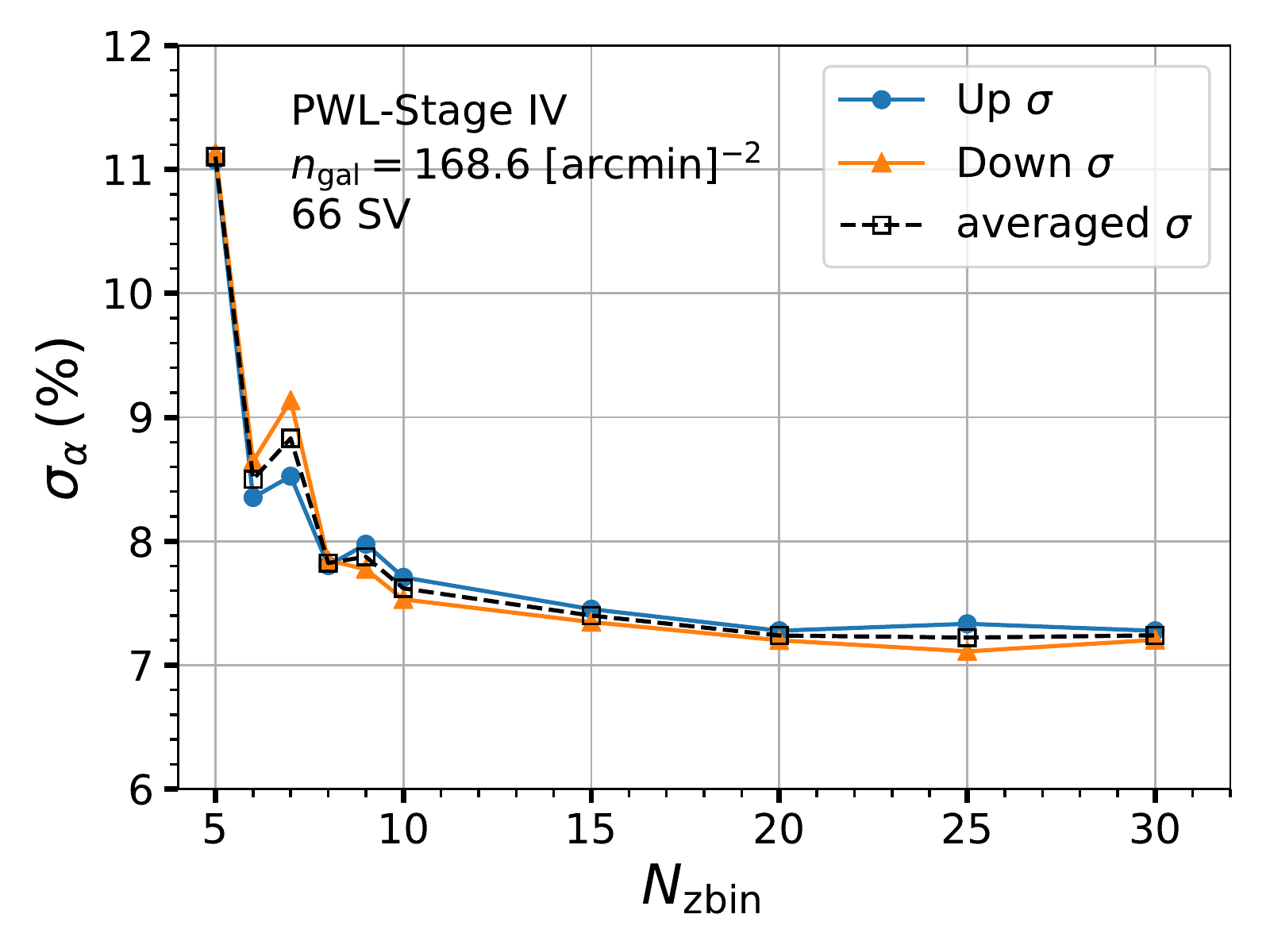}
\includegraphics[width=0.45\linewidth]
{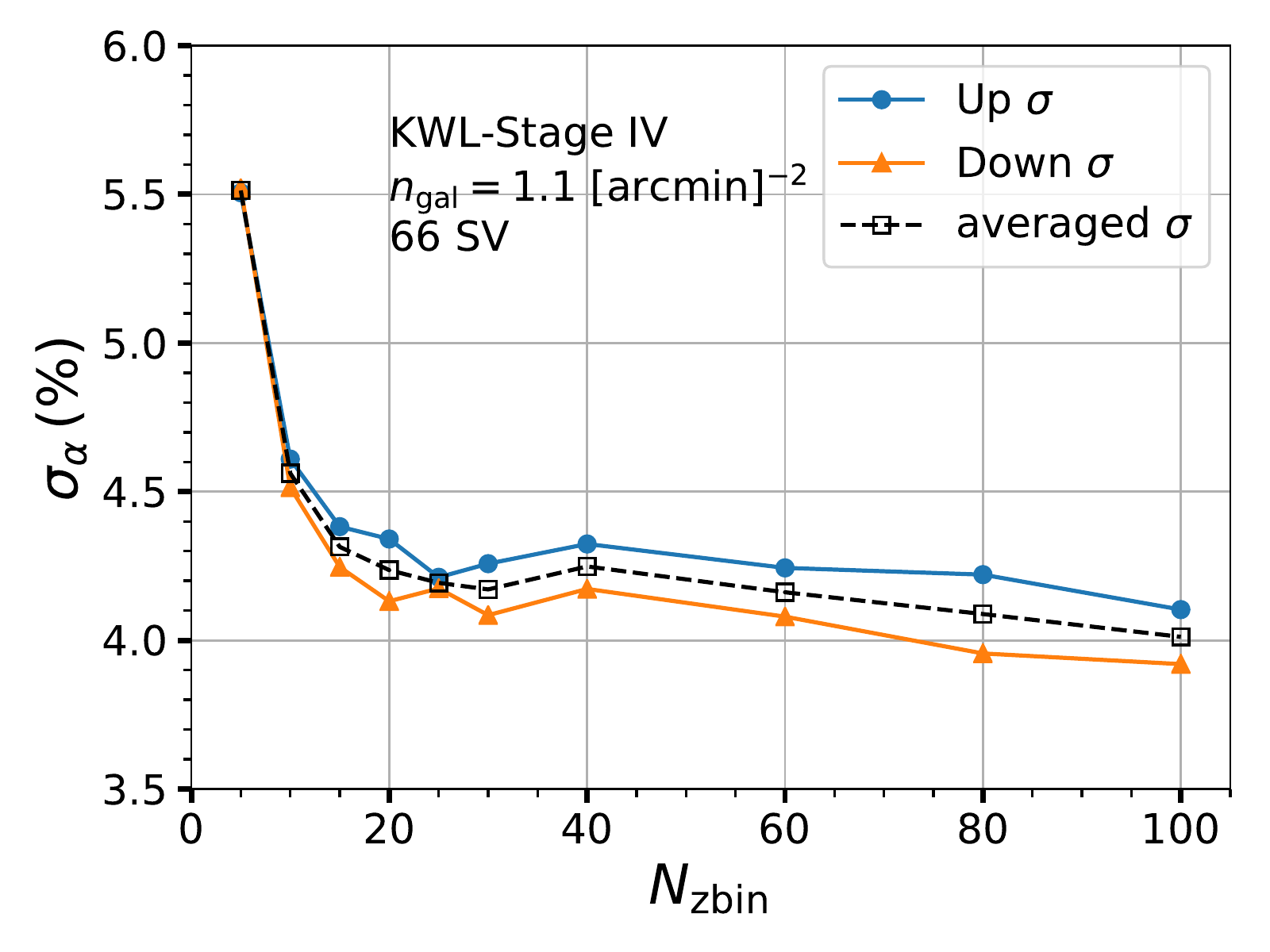}
\caption{The BAO constraints as a function of the redshift bin width: the low noise PWL-Stage IV (left panel), and the KWL-Stage IV (right panel). We use $\Nmodes= 66$. In each panel, circular and triangular points denote upper and lower $\sigal$ from our MCMC fitting, respectively, and empty squares with a dashed line are the average of the two $\sigal$. We find that both surveys show convergence near $\Nzbin=30$.}\label{fig:Nzbineffect}
\end{figure*}

\begin{figure}
\includegraphics[width=0.95\linewidth]
{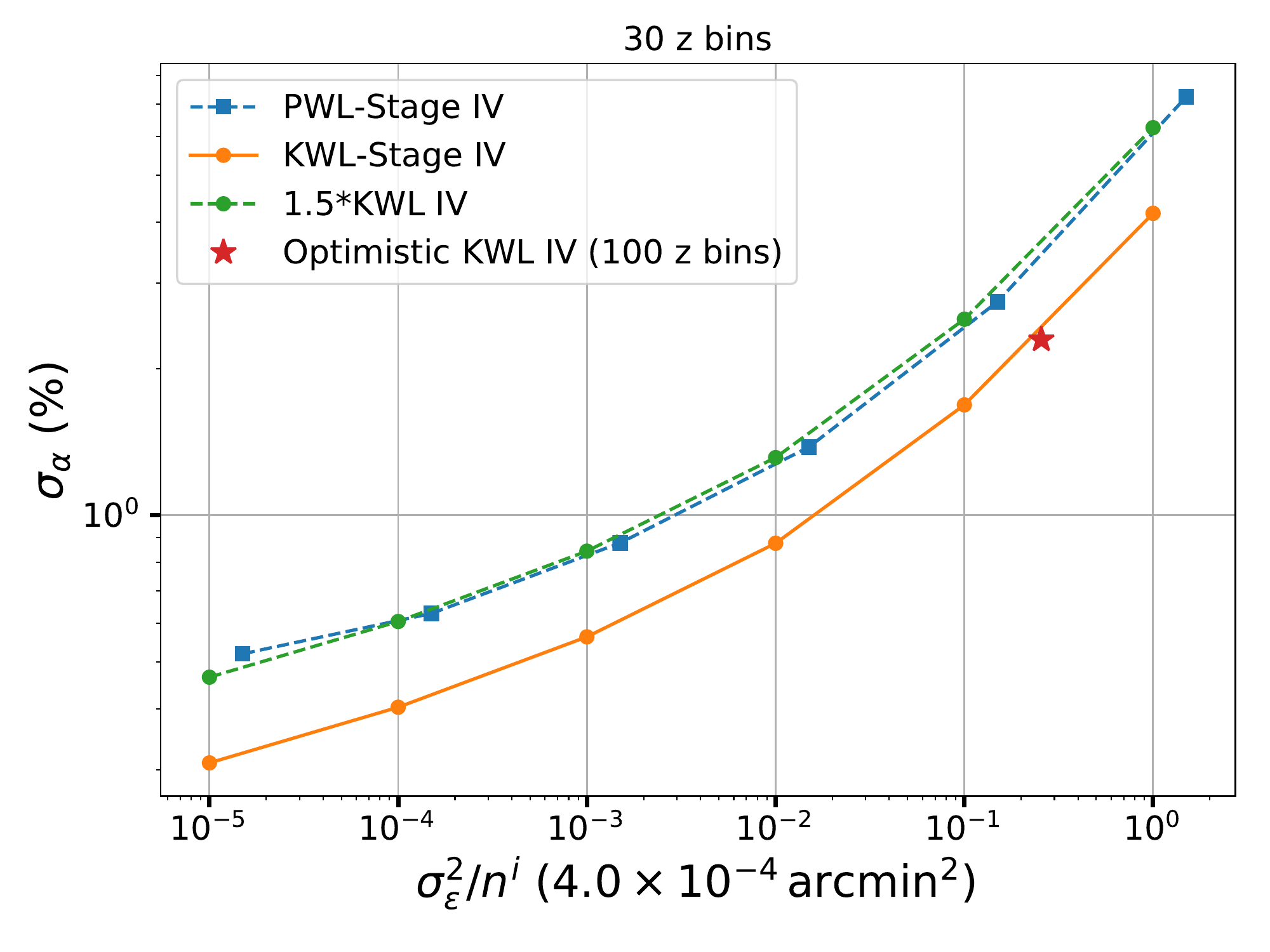}
\caption{The BAO scale constraint as we decrease the shape noise further below the default values. The x-axis shows the factor to multiply to the default shape noise of the KWL-Stage IV. I.e., the rightmost points correspond to the default KWL-Stage IV (orange circles) and the low noise PWL-Stage IV (blue squares). The PWL-IV points are systematically shifted to the right to account for its effective signal to noise that is lower by $\sqrt{1.4}$ than the KWL-Stage IV. Both results use $\Nzbin=30$. The green circles are the rescaled KWL-Stage IV constraints by a factor of 1.5 to show that the multiplicative offset between the PWL and the KWL surveys remain constant. The single star point corresponds to the optimistic KWL-Stage IV with $\Nzbin=100$ that will be discussed in \S~\ref{subsec:optKWIV}. The point is very close to the curve for $\Nzbin\sim 30$ at the same noise level, which implies that the information within the KWL survey is mostly extracted with  $\Nzbin\sim 30$ even with lower noise.}\label{fig:shapenoise}
\end{figure}

The error bars reflect the typical band power fluctuations we expect for each case. 
Again, neighboring $k$ bins are highly correlated, more so as we increase $\Nmodes$, such that the diagonal errors alone do not fully reflect the correlated fluctuations. In the lower sub-panels, we show the cross correlation coefficients $r_{ij}$ between $k$ bins~\footnote{$r_{ij} = \frac{\text{Cov}[P'_{\delta, \, i},\ P'_{\delta,\, j}]}{\sqrt{\text{Cov}[P'_{\delta, \, i},\ P'_{\delta,\, i}] \times \text{Cov}[P'_{\delta, \, j},\ P'_{\delta,\, j}]}}$
} with $k_i=0.03$, $0.1$ and $0.2$ $\ihMpc$. Each panel quotes $\Nmodes$ we choose for each figure, which is approximately the minimum number of unreplaced SV modes reaching the convergence in the BAO detection level, the reduced $\chi^2$, and $\sigma_\alpha$.

Although misleading without taking the off-diagonal covariance into the account, the large diagonal errors in the case of PWL-Stage III (top left) in Fig.~\ref{fig:fit_Pwnw} forewarn a difficulty retrieving BAO scale information from the cosmic shear of a DES-like survey. Indeed, $\sigma_\alpha$ for the PWL Stage III and IV are 33\% and 22\%, respectively, implying no meaningful BAO constraint even with the Stage IV case. As a comparison, \citet{Simpson2006} predicted a $2-\sigma$ detection of the BAO from a PWL-Stage IV. The two results are fairly consistent and the difference could be partly due to the number density and the redshift precision that are better than assumed here, but is probably because of their $\ell$-dependent redshift binning of source galaxies to manipulate the lensing kernel to reduce the line-of-sight projection of the BAO feature.

On the other hand, the KWL surveys are more effective in recovering the BAO feature.
While the KWL-Stage III would not be able to return a meaningful BAO constraint (i.e., $\sigma_\alpha=16\%$), the KWL-Stage IV gives $\sigma_\alpha = 4.1\%$ likely due to the larger volume and the smaller BAO damping expected at the higher median redshift, which therefore could be used for a consistency check against the percentage level BAO constraint from the current galaxy surveys \citep[e.g.][]{Alam16}.
Fig.~\ref{fig:fit_Pwnw} also quotes the significance of BAO detection for these  surveys. 
As expected, the PWL Stage surveys cannot detect BAO while KWL Stage III can give about a $1-\sigma$ detection. Meanwhile, the KWL-Stage IV predicts the BAO detection at $3.1-\sigma$. The precision and the detection of the KWL Stage IV would therefore correspond to the first galaxy BAO detection by \citet{Eisenstein05}.

\subsection{Understanding the differences between the KWL and PWL surveys}\label{subsec:dependence}

The KWL surveys have advantages against the PWL surveys in the redshift precision and shape noise. In this section we test the effect of such advantages.
\subsubsection{The effect of the shape noise difference}
We first test the effect of the shape noise by matching the shape noise of the PWL-Stage IV to that of the KWL-Stage IV. Fig.~\ref{fig:PW_lowsn} shows that if we decrease the shape noise level of the PWL-Stage IV to that of the KWL-Stage IV (hereafter `low-noise PWL-Stage IV'),  $\sigma_\alpha$ decreases from 22\% to 7\% and the detection level increases to $2-\sigma$. This is still larger than $\sigma_\alpha = 4\%$ from the KWL-Stage IV. As mentioned earlier, the amplitude of $C_\ell$ of the PWL-Stage IV tends to be lower than that of the KWL-Stage IV of the same source redshift bin, except for that at lower redshifts, due to its lower lensing kernel (in Fig.~\ref{fig:leneff}). Indeed if we compare the cumulative signal to noise squared (i.e., $\sum (S/N)^2$)~\footnote{$\hat{C}^T \mathbb{C}^{-1} \hat{C}$ where  the covariance matrix $\mathbb{C}$ is from Eq.~(\ref{eq:cov_shear})} of $C_\ell$ for all $\ell$ within the mock data range and redshift bins, the KWL-Stage IV has 1.4 times more $\sum (S/N)^2$ than the PWL-Stage IV.  Therefore, the low noise PWL-Stage IV would have 5.9\% of the BAO constraint if its signal to noise were more precisely matched to the KWL-Stage IV.
This still leaves a factor of 1.43 offset between the constraints of the two surveys of the same signal to noise, indicating that the shape noise relative to the amplitude is not the sole major advantage of the KWL survey.

\subsubsection{The effect of $\Nzbin$}\label{subsec:Nzbin}
Before looking for other advantages, we first make sure that this remaining difference is not due to the different choices of tomographic bins as we use $\Nzbin = 30$ for the PWL surveys and $\Nzbin=100$ for the KWL surveys. Fig.~\ref{fig:Nzbineffect} shows the $\sigma_\alpha$ using all $\Nmodes$ for a different choice of $\Nzbin$. We find that $\sigma_\alpha$ of the KWL-Stage IV does not increase much when decreasing $\Nzbin$ from 100 to 30 (in the right panel),
which implies that the difference remains even with the same redshift binning. For both the low-noise PWL-Stage IV and the default KWL-Stage IV surveys, $\sigma_\alpha$ increases if we use redshift bins coarser than $\Nzbin = 30$.

The slow improvement with increasing $\Nzbin$ from 30 (i.e., $dz < 0.07$) to 100 in fact implies that we are not taking the full advantage of the redshift resolution of the KWL surveys, possibly because the effect of the lensing efficiency kernel is not fully de-convolved in the presence of the default noise. It could be also possibly because the Limber approximation we used limits our extracting the line-of-sight Fourier modes; in Appendix \S~\ref{sec:limber}, we estimate that the effect of the Limber approximation and find that it is unsubstantial.

If the lensing kernel of the KWL survey is not deconvolved enough to take advantage of its small redshift error, and also if the remaining difference between the low-noise PWL-Stage IV and the KWL-Stage IV is due to the greater redshift error and the corresponding lensing kernel of the former that is perhaps more difficult to be deconvolved even in the presence of the similar shape noise, we expect that the difference in $\sigal$ of the two surveys would decrease as we further decrease the assumed shape noise. This is because we should be able to deconvolve both kernels at the zero noise limit with the perfect data.
Fig.~\ref{fig:shapenoise} shows $\sigma_\alpha$ of the PWL and KWL surveys, with $\Nzbin=30$, as we decrease the shape noise. Even with the smallest shape noise we could numerically test in this paper, the ratio between the KWL survey and the PWL survey remains almost constant. We believe that, as shown in Fig.~\ref{fig:Ciil_shapenoise}, the lensing kernels of PWL and KWL surveys are both very broad such that the distinction between the kernels of different source redshift bins within each survey is too small to fully deproject the density fluctuations along the line of sight even in the presence of the smallest noise we could test.

\subsubsection{The effect of the lensing kernel shape}
\begin{figure}
\includegraphics[width=0.9\linewidth]
{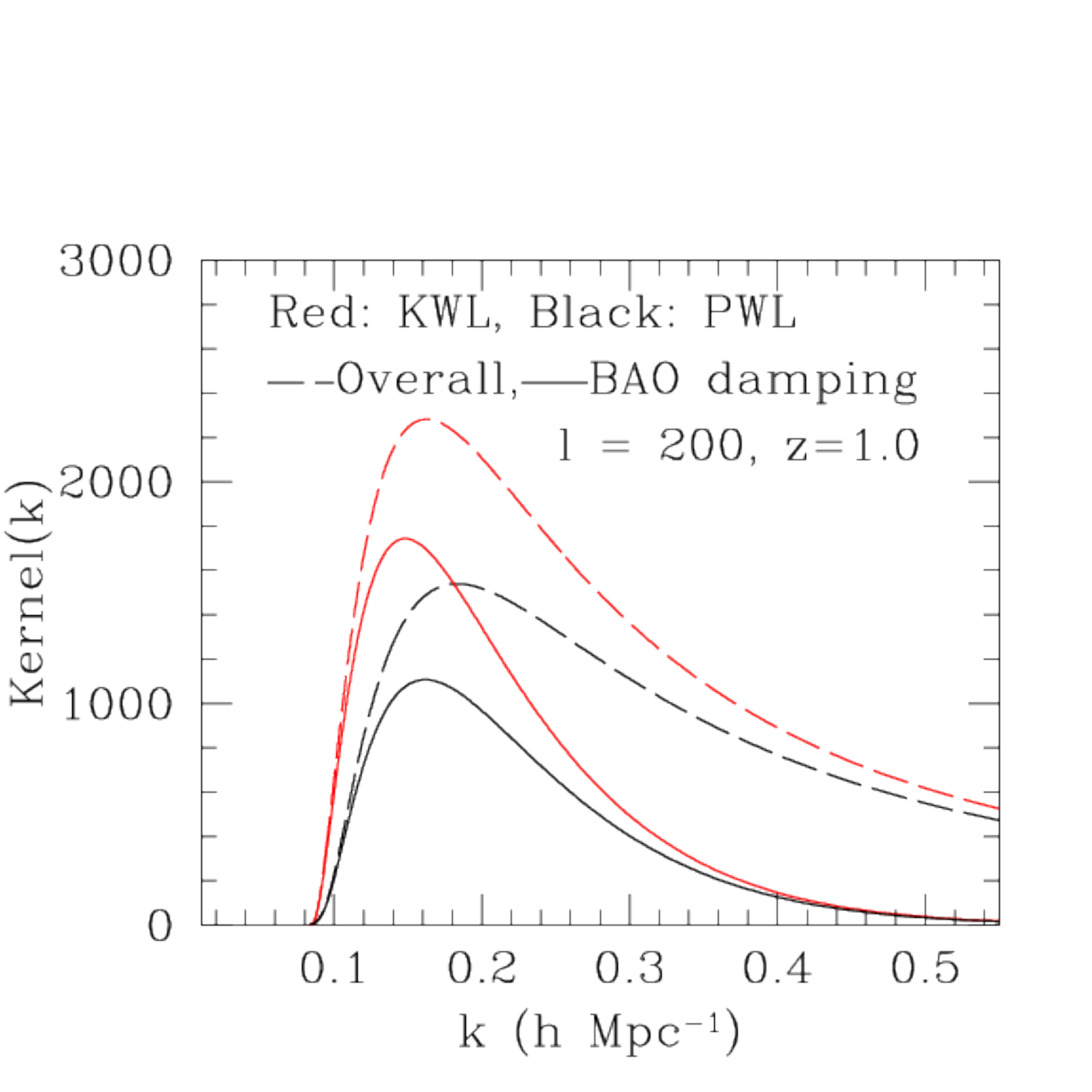}
\caption{The lensing window function that convolves the underlying matter power spectrum in Eq.~\ref{eq:CijlKW}. For $C(\ell=200)$ and the $z=1-1.065$ source redshift bin. Dashed lines: the window function that convolves the overall power spectrum. Solid lines: the window function multiplied with the BAO damping factor $\exp{(-k^2\Sigma^2/2)}$ to represent the scales over which the BAO information mainly resides.}\label{fig:kernelk}
\end{figure}

We believe that the remaining difference between the PWL and the KWL surveys, when the signal to noise of the amplitude is matched, is related to the more distinct BAO feature of the KWL survey as shown in the bottom panels of Fig.~\ref{fig:Ciil_shapenoise}. Note that the BAO feature in the input power spectrum is the same for both cases. Then, one can question why the KWL survey is more sensitive to the BAO feature when its kernel is as broad as the PWL survey. It is likely because of the shape of the lensing kernel. The KWL lensing kernel in Fig.~\ref{fig:leneff} is above the PWL lensing kernel at high redshift and as a result, the KWL survey weighs lower $k$ information that is contributed from higher redshift clustering. Figure \ref{fig:kernelk} shows the corresponding lensing window function that convolves the underlying matter power spectrum in Eq.\ref{eq:CijlKW} for $C(\ell=200)$ for the $z=1-1.065$ source redshift bin, as an example. The dashed lines are the window function that convolves the overall power spectrum. The solid lines are the window function multiplied with the BAO damping factor $\exp{(-k^2\Sigma^2/2)}$ to represent the scales over which the BAO information mainly resides.
The ratio of the area covered by the solid line (i.e., more relevant to the BAO) v.s. the dashed line (i.e., relevant to the overall power spectrum) upto $k= 2.43\ihMpc$ is indeed greater for the KWL survey by $
\sim 17\%$ for the case presented, which is qualitatively consistent with Fig.~\ref{fig:Ciil_shapenoise}. That is, the BAO contribution would be greater for the KWL survey because its lensing kernel weighs high redshift (i.e., lower $k$ that are more relevant for the BAO) more than the PWL survey does, as a result of the source redshift distribution of the former being almost the delta function. When the observed $C_\ell$ data over $\ell \le 2002$ was adopted and deprojected to $P(k)$ band powers, the deprojected $P(k)$ of the KWL surveys is then  more sensitive to the BAO feature.

Following this reasoning, we recalculated the cumulative signal to noise of the reconstructed power spectrum $P(k)$. When all $k$ was included, the ratio of $\sum (S/N)^2$ between the two surveys was again $\sim 1.4$, which is consistent with the $\sum (S/N)^2$ from $C_\ell$. However when we limited the range to be the BAO fitting range, i.e., $0.015<k<0.3\ihMpc$, the ratio of $\sum (S/N)^2$ increased to 1.7. Therefore the signal to noise of the amplitude of the power spectrum over $0.015<k<0.3\ihMpc$ is responsible for a factor of 1.3 difference between the low noise PWL-Stage IV and the KWL-Stage IV BAO constraints and the remaining 1.3 difference is, we believe, largely due to the different degree of the BAO information in the two surveys. In order to further identify where the extra BAO information resides, we repeated the $\chi^2$ fitting to the reconstructed $P(k)$ band powers of the low noise PWL IV using the covariance matrix of the KWL IV survey. Just by replacing the covariance matrix, we recovered almost the same result as the KWL IV survey constraints: i.e., $\sigal=4.2\%$ and $3.1\sigma$ of the detection level. That is, the KWL survey not only has a greater $S/N$ for the amplitude of the power spectrum, but also a greater BAO information stored in the covariance of the power spectrum.

To summarize, we were not able to fully deconvolve the lensing kernel of the KWL survey even with the smallest shape noise we could test and we believe it is because the typical kernel is wide and the difference between the kernels of neighboring source redshift bins is very small to be used to deproject the clustering information along the line of sight. This implies that the recovered information from the KWL survey will be subject to the line-of-sight projection that is similar to the photometric galaxy surveys rather than the spectroscopic galaxy surveys. This also explains why the Limber approximation we used for the KWL surveys did not affect our final BAO constraint much (in Appendix \S~\ref{sec:limber}). Despite the inefficient deprojection, we find that KWL surveys have advantages for the BAO measurements in terms of the smaller shape noise we can achieve with the spectroscopic data, the lensing kernel that weighs larger scale clustering from higher redshifts, and the resulting covariance structure. The factor of improvement we find is about 1.7 on the BAO scale when the shape noise to the source galaxy number was naively matched. 

\subsection{Improving the KWL-Stage IV and Discussions}\label{subsec:optKWIV}

Encouraged by the BAO constraint from the KWL-Stage IV survey, we explore the precision improvement with decreasing shape noise further from the default KWL- Stage IV. We find that an optimistic scenario for the KWL-Sage IV with $n_{\rm gal}=4.3/{\rm arcmin}^2$, i.e., with the shape noise four times lower than the default value, returns $\sigma_\alpha \sim 2.3\%$ (shown in Fig.~\ref{fig:shapenoise} as a star point) with $4.9-\sigma$ of the BAO detection, which is about two times better than the default KWL-Stage IV BAO constraint.

The 2\% BAO constraint at $z\sim 1$ is comparable to a typical BAO constraint from an effective volume of $\sim 0.3\trihGpc$ of a galaxy survey at this redshift with the BAO reconstruction technique~\citep{ESSS}. For example, \cite{Bautista17} reports 2.8\% using luminous red galaxies in the SDSS IV Data Release 14 of $0.9 {\rm Gpc}^3$. \cite{FR14} predicts 0.6\% for 4 $\trihGpc$ at $z\sim 1$ after applying the BAO reconstruction technique. 
Here we attempt to explain why the BAO constraint from the KWL-Stage IV is still much less precise than what is typically expected from a galaxy clustering data of the source galaxies given the effective volume~\footnote{The effective volume would depend on the details of the FKP weight~\citep{FKP}.} of  $\sim 100\Gpc^3$ with the maximum comoving galaxy number density of $0.002\itrihMpc$ at $z\sim 0.7$. From Fig.~\ref{fig:Ciil_shapenoise}, one estimates the signal to shape noise per mode at $\ell\sim 200$ and at  $z\sim 0.7-1$ for the KWL-Stage IV case is approximately 10 times less than the signal to shot noise of the angular clustering of galaxies at $\ell\sim 200$ \citep[e.g.,][]{Ho2012} (i.e., $\nbar C_{\ell,{\rm galaxy}}/(\nbar C_{\ell,{\rm shear}}/\sigma^2_\epsilon)$), assuming the galaxy number density is the same as the source galaxy number density for the KWL-Stage IV. A $1\trihGpc$ galaxy survey with $\nbar=0.002\itrihMpc$ galaxies is expected to return a BAO constraint of $\sim 1.3\%$ without the density field reconstruction near this redshift. A survey with 10 times greater shot noise (assuming no redshift-space distortions as in KWL surveys) will return 3.4\% of the constraint. As we argue in \S~\ref{subsec:dependence}, we believe that the KWL-Stage IV survey is still subject to a considerable line of sight smearing, which would degrade the BAO precision just as if a photometric galaxy survey is subject to such degradation \citep{Seo2003}.  If we assume the photometric error that corresponds to the redshift width of $\Nzbin=30$ (i.e., $\Delta \chi = 88\hMpc$), we expect a BAO constraint of about 12\%. Therefore, we expect at least a factor of $\sim 10$ worse BAO constraint from the KWL-Stage IV compared to the pre-reconstruction galaxy BAO survey. Assuming reconstruction improves the galaxy BAO by a factor of 2, the total factor of 20 roughly explains the discrepancy between our study and what is expected from a galaxy survey given the effective volume.

Nevertheless, although the naive precision comparison demonstrates that even the optimistic KWL-Stage IV under-performs compared to the galaxy BAO surveys, the BAO constraint we studied in this paper is an {\it additional gain} to the information content of the KWL surveys reported in Huff13. For example, Huff13 predicts the information content of the default KWL-Stage IV, without the BAO information, to be 7 times that of the Dark Energy Survey.

Another significance of the KWL-Stage IV survey would be that this method would provide the first (and the only) detection of the BAO signature directly from late-time matter at $\sim 5\sigma$ significance with $\sim 2\%$ precision. The BAO constraint from matter can be compared to the BAO constraints from galaxy surveys to detect a potential BAO scale bias induced in the galaxy surveys due to the supersonic streaming velocity of baryons at high redshift \citep{Dalal10, Tseliakhovich10, Blazek16, Slepian15, Schmidt16, Schmidt17}. Although it is not likely that such bias is as large as 2\%, as \cite{Blazek16} for example estimates $\sim 0.5\%$ of the BAO scale shift by $\sim 1\%$ of density fluctuation by the streaming velocity bias, we expect that the cross-correlation between the cosmic shear and the galaxy density field within the KWL survey can potentially return a much greater precision on the BAO scale bias due to the sample variance cancellation between the two tracers within the same cosmic volume, given that the cosmic shear information alone returns 2\% of the BAO constraint. We plan to investigate the BAO in cross-correlations between galaxies and shear within the KWL surveys in future studies. Also, knowing that the lensing kernel difference affects the BAO precision, we could optimize the source galaxy redshift binning of the KWL surveys for the BAO constraint, benchmarking the method in \citet{Simpson2006}. We leave this for future studies as well.

\section{Conclusions}\label{sec:con}
In this paper, we investigated the feasibility of directly detecting the BAO feature in the dark matter distribution using spectroscopic, kinematic weak lensing (KWL) surveys proposed by \citet[][]{Blain02,Morales06,Huff13}. We simulated cosmic shear tomography analyses of future KWL surveys. We extracted the spacial power spectrum from the simulated shear angular power spectrum data,  isolated the BAO feature, and constrained the BAO scale parameters for different survey conditions. We compared such results with the Stage III and Stage IV photometric weak lensing (PWL) surveys. We publicize the code used for this paper in \url{https://github.com/zdplayground/SVD_ps}. We summarize our findings below.

As we expected, the PWL surveys could not constrain the BAO scale due to the large shape noise and the large redshift uncertainty. Meanwhile we found that we can extract the BAO information from a KWL-Stage IV survey and potentially derive a $\sim 4\%$ constraint on the BAO scale. A more optimistic assumption on KWL-Stage IV predicts a BAO detection at $4.9\sigma$ significance and a $\sim 2\%$ constraint on the BAO scale. 

We found that both the lower shape noise and the shape of the lensing efficiency kernel that weighs higher redshift and lower $k$ information are responsible for the BAO constraint from the KWL surveys. It appears that the BAO information is effectively encoded in the covariance between different $k$ scales. In detail, our analysis implies that we do not take full advantage of the redshift precision of the KWL surveys probably because the lensing efficiency kernel is very broad in redshift and the level of the shape noise, although much smaller than the PWL surveys, does not permit a full de-convolution of such broad lensing-efficiency kernel. That is, the clustering information from the KWL surveys will be subject to the line of sight projection that is more comparable to the photometric galaxy clustering data rather than to the spectroscopic galaxy clustering data. 

While a realization of such KWL surveys would be based also on cost-wise consideration, we discuss the scientific significance of such measurements. 
First, this method can provide the first (and the only) detection of the BAO signature directly from late-time matter, which is mostly dark matter. The BAO constraints from dark matter and from galaxy surveys can be compared each other to constrain a potential BAO scale bias induced in the galaxy surveys due to the supersonic streaming velocity of baryons at high redshift 
\citep{Dalal10, Tseliakhovich10, Blazek16, Slepian15, Schmidt16, Schmidt17}. 
Of course, the 2\% BAO constraint from dark matter at $z\sim 1$ even by the optimistic KWL-Stage IV is not small enough to be compared to the sub-percentage BAO constraint from the galaxy surveys. However, we believe that our prediction is likely a lower limit estimate of the BAO information available. Such spectroscopic survey will by default allow measurements of 2-point statistics of galaxy density field as well as the cross-correlation between the spectroscopic galaxy density field and the cosmic shear field.  The cross-correlation within the same volume can potentially return a much greater precision on the BAO scale bias due to the sample variance cancellation between the two large-scale structure tracers, i.e., dark matter and the galaxies. We plan to extend our study to include the BAO information from such cross-correlations in future. We also plan to investigate for optimization of the lensing kernel for the BAO constraint, by manipulating the source galaxy distribution of each tomographic bin, similar to what is done in \citet{Simpson2006}.

We note that our study identified additional gain to the information content of the KWL surveys reported in Huff13. Huff13 predicts the cosmological information of cosmic shear tomography assuming the default KWL-Stage IV, without the BAO information, would be 7 times that of the Dark Energy Survey. 

There are a few theoretical as well as observational complications we bypassed in this analysis. Among many, we note that we did not include the nonlinear evolution effect on the overall shape of cosmic shear power spectrum by means of halofit \citep[e.g.][]{Smith03, Takahashi12}. Since we focus on the BAO information alone, we believe that our simplification is warranted. Instead we included the nonlinear smearing of the BAO feature at the characteristic redshift of the surveys. We also did not include the non-linear (non-Gaussian) covariance on small scales due to structure formation. While such effect will degrade the BAO constraint further,  we do not expect it to be significant. Projections along the line of sight is known to reduce non-Gaussianity in the cosmic shear signal (simply due to the central limit theorem argument) and the degradation in the final cosmological parameters due to the remaining non-Gaussianity is shown to be even smaller than what is expected from the degradation in the signal-to-noise of the amplitude~\citep[e.g.][]{Takada09}. Our default KWL surveys still largely suffer the projection effect as we show in this paper, and in addition, the BAO feature we investigate in this paper is on a much larger scale ($\ell$ effectively less than 1000) where non-Gaussianity is less significant than a typical cosmic shear information. Even without the projection effect, \citet{Ngan12} reports less than a 15\% discrepancy in the BAO constraint due to the Gaussian approximation of the covariance matrix. On the other hand, the super-sample covariance effect we did not include could be non-negligible for our forecasts ~\citep{BarrieraSSC}. We plan to include such details in future study while this paper is a pilot study that motivates such extended efforts. 

\section*{Acknowledgements}
We greatly appreciate valuable comments from Christopher M. Hirata and Patrick Mcdonald. We also thank Tim Eifler for providing the source galaxy redshift information for the PWL-Stage III and IV surveys adopted in this work.
Z.D., D.C., and H.-J.S.~are supported by the U.S.~Department of Energy, Office of Science, Office of High Energy Physics under Award Number DE-SC0014329. 
S.S. was supported in part by JSPS KAKENHI Grant Number JP15H05896 and by World Premier International Research Center Initiative (WPI Initiative), MEXT, Japan. 
 This research used the Dark Energy Spectroscopic Instrument (DESI) allocation resources of the National Energy Research Scientific Computing Center (NERSC), a U.S. Department of Energy Office of Science User Facility operated under Contract No. DE-AC02-05CH11231.

\appendix
\section{The effect of the Limber approximation}~\label{sec:limber}
\begin{figure}
\includegraphics[width=0.9\linewidth]
{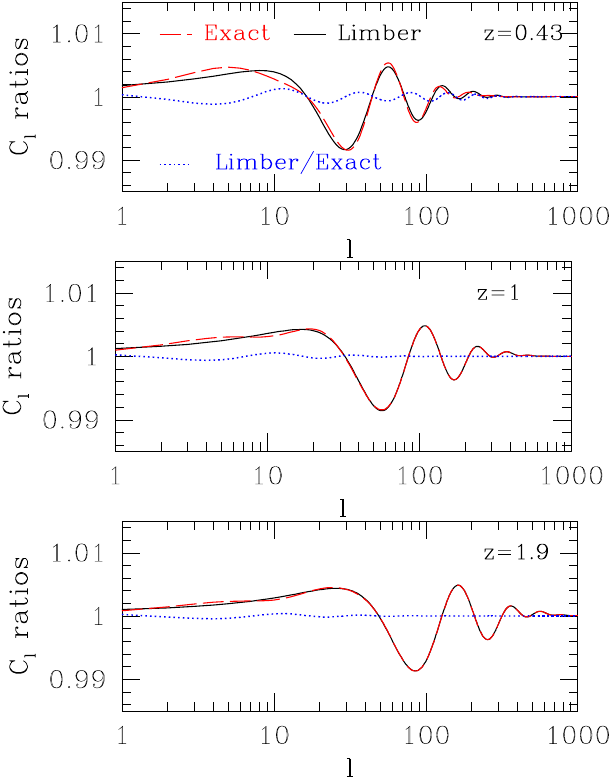}
\caption{The effect of the Limber approximation. The ratios between the $\Pwig/\Pnow$ of the auto power spectra from the Limber approximation and from the exact calculation. At three source redshift bins of the KWL-Stage IV survey, at $z=0.43$, 1, and 1.9; while the exact calculation returns slightly more distinct BAO feature, especially at $z=0.43$, the difference is very small except for low $\ell$ modes. The difference on the overall shape is removed in this figure.}\label{fig:limbervsexact}
\end{figure}

We qualitatively estimate the effect of the Limber approximation for the KWL survey in Eq.~\ref{eq:CijlKW}. Fig.~\ref{fig:limbervsexact} shows the auto angular power spectrum at three source redshift bins of the KWL-Stage IV survey, at $z=0.43$, 1, and 1.9; while the exact calculation returns slightly more distinct BAO feature, especially at $z=0.43$, the difference is very small except for low $\ell$ modes. The overall amplitude of $C(\ell)$ would be also offset only by a few percent for $\ell < 10$, which should not affect our results much. We observe a similar level of discrepancy in the cross power spectra. The good agreement between the Limber approximation and the exact calculation is expected since the lensing kernel of the KWL survey is very broad. Without a full calculation, we estimate that the Limber approximation did not substantially affect our main results.

\end{document}